\documentclass[useAMS,usenatbib]{mn2e}
\usepackage{amssymb}
\usepackage{graphicx}

\newcommand{\dg}{^{\rm{}o}}

\newcommand{\rs}{R_{\mathrm{S}}}
\newcommand{\fvec}[1]{\mbox{\protect\boldmath$#1$}}
\newcommand{\rph}{r_{\rm{}ph}}

\begin{document}

\title[On the role of strong gravity in polarization]{On
 the role of strong gravity in polarization from scattering 
 of light in relativistic flows}
\author[J.~Hor\'ak and V.~Karas]{J.~Hor\'ak and V.~Karas\\~\\
Astronomical Institute, Academy of Sciences,
 Bo\v{c}n\'{\i}~II, CZ-140\,31~Prague, Czech Republic\\
Charles University, Faculty of Mathematics and Physics,
 V~Hole\v{s}ovi\v{c}k\'ach~2, CZ-180\,00~Prague, Czech Republic}

\date{Accepted ............ 2005; Received ............. 2005}
\pagerange{\pageref{firstpage}--\pageref{lastpage}}
\pubyear{2005}
\maketitle
\label{firstpage}
\begin{abstract} We study linear polarization due to scattering of light
on a cloudlet of particles, taking into account the radiation drag and
the gravitational pull exerted on them by a central body. Effects of
special and general relativity are included by connecting a model of
Beloborodov (1998) for the local polarization of scattered light with
Abramowicz, Ellis \& Lanza (1990) formalism for  the particle motion
near an ultra-compact star. Compactness of the central body and its
luminosity  are two critical parameters of the model. We discuss the
polarization magnitude of photons, which are Thomson-scattered into
direct and higher-order images. Importance of the latter is only
moderate under typical conditions, but they may give rise to distinct
features, which we explore in terms of a toy model. The scattered signal
exhibits variations of intensity and of polarization with mutual
time-lags depending on the beaming/focusing effects and the light travel
time.
\end{abstract}


%-------------------------------------------------------------------
%  Introduction
%-------------------------------------------------------------------

\section{Introduction}
Scattering of ambient light by fast moving flows has been
recognized as a conceivable mechanism operating in different classes of
objects. This process quite likely contributes to the linear
polarization of initially unpolarized soft radiation up-scattered in
blazar jets \citep{beg87}, winds from accretion discs \citep{bel98} and
in gamma-ray bursts (\citealt{sha95}; \citealt{laz04}; \citealt{lev04}).
A significant level of intrinsic linear polarization of $\Pi\sim3$\% was
reported in a microquasar GRO J1655-40 \citep{sca97} and similarly for
LS\,5039 \citep{com04} in the optical band. A model of magnetized
fireball polarization was discussed by \citet{ghi99}, who demonstrate
that the expected polarization lightcurve should exhibit two peaks.
Variations of the Galactic Centre linear polarization were reported in
the millimetre band \citep{bow05}. Recently, \citet{vii04} brought
the attention to strong-gravity effects in polarimetry of accreting
millisecond pulsars.

The idea of X-ray polarization studies providing clues to the physics of
accreting compact objects was discussed in seminal papers (\citealt{ang69};
\citealt{bon70}; \citealt{lig75}; \citealt{ree75}; \citealt{sun85}).
Here we study a simple model with Thomson scattering on electrons 
ejected outwards from the centre or falling back;
a novel point is that we consider strong gravity effects. The source of
seed photons can be identified with the surface of a central star lying
in arbitrary (finite) distance from the scatterer. Alternatively, it can
represent an axially symmetric accretion disc near a black hole, a
quasi-isotropic (ambient) radiation field, or a combination of all these
possibilities. In order to avoid numerical complexities we do not
discuss other radiation mechanisms that also produce polarization,
namely, we do not consider synchrotron self-Compton process, which is the most
likely process wherever magnetic fields interact with relativistic
particles (see \citealt{pou94}; \citealt{cel94} for discussion and 
for further references).

In the non-relativistic regime, a conceptually similar problem was
examined by \citet{rud78} and \citet{fox94}, who considered polarization
of light of a finite-size star due to scattering on free electrons in a
fully ionized circumstellar shell. These works showed non-negligible
solid angle of the source subtended on the local sky of scattering
particles has a depolarizing influence on the observed fractional
polarization. This result was discussed by various authors, because the
scattering of stellar continuum is a probable source of net polarization
of visible light in early-type stars, for which polarimetry has become a
standard observational tool (e.g.\ \citealt{poe76};
\citealt{bro77}). General relativity effects are negligible in these
objects, however, even the Newtonian limit is complex enough: the
observed signal does not easily allow to disentangle the contribution of
a circumstellar shell, the primary radiation of the star and the effect
of interstellar medium. In our case the situation complicates further,
because the system is highly time-dependent and mutual time delays along 
different light rays must be taken into account. 

The general relativity signatures, which we discuss in this paper, 
reach maximum in a system containing a compact body with radius less
than the  photon circular orbit, i.e.\ $R_\star<\rph$. This can arise
either as a result of a non-stationary system with
$R_\star{\equiv}R_\star(t)$, presumably during a gravitational collapse,
or it can represent a static ultra-compact star if such exist in nature.
Here we impose spherical symmetry of the gravitational field, and so
both situations are identical as far as the form of spacetime metric is
concerned. The Schwarzschild vacuum solution describes the external
gravitational field of all types of compact objects within general
relativity, provided that their rotation is negligible and self-gravity
of accreting matter does not contribute significantly to the
gravitational field. We will assume that these constraints are
fulfilled. A sequence of $R_\star=\mbox{const}$ situations can be 
employed in order to model a collapsing case.

The formalism that we apply works for a system with arbitrary
compactness, even if the case of  $R_\star=\mbox{const}<\rph$ seems to
be an unrealistically large compactness per se, likely violating the causality
condition for neutron stars. Such a high compactness would thus be normally
excluded, however, the situation has not yet been definitively settled
(see \citealt{lat04} for a recent review). According to astrophysically
realistic equations of state, neutron-star sizes do not reach
ultra-compact dimensions; their typical radii should exceed the photon
circular orbit (\citealt{lat01}; \citealt{hae03}), and so the
gravitational effects are constrained accordingly. However, the 
possibility of $R_\star$ being slightly less than $\rph$ persists, and
there seems to be a growing awareness now of the fact that  the very
high density regime needs to be explored further. 

The interest in ultra-compact stars has been recently revived mainly in
connection with gravitational waves in general relativity 
(\citealt{cha91}; \citealt{kok04}). One may consider also more exotic
options. Strange quark stars (e.g.\ \citealt{alc86}; \citealt{dey98})
can have their radii extending down to the Buchdal limit (see
\citealt{web05} for a recent review on other forms of quark matter and
their relevance for compact stars). If nucleons can be confined at the
density lower than nuclear matter density, then Q-stars could exist with
a relatively high mass of $\sim10^2M_\odot$ and the radius as small as
$R_\star\sim0.9\rph$ (e.g.\ \citealt{mil98}). On a more speculative
level is the idea of gravastars \citep{maz04}, compactness of which can
exceed the Buchdal limit. Options for detecting the thermal radiation
from different kinds of ultra-compact stars have been recently discussed
by \citet{mcc04}. In the case of transiently accreting neutron stars,
the crustal heating has been nominated as one of relevant mechanisms
generating thermal emission from the surface (\citealt{hae90};
\citealt{bro98}). 

The issue of realistic equations of state for ultra-compact star matter
is beyond the scope of the present paper, likewise the actual
observational information on masses and radii of compact stars (see e.g.
\citealt{hae03}). It will be convenient to introduce a dimensionless
parameter, $\zeta\equiv1-R_\star/r$, which maps the whole range of radii
above the star surface onto $\langle0,1\rangle$ interval, so that the
form of graphs does not change with the star compactness. We consider
the whole range of $0\leq\zeta\leq1$ allowed by general relativity. It
is worth noticing that the formalism used below could be readily applied
also to the case of an accretion disc as a source of light near a black
hole. Obviously, a black hole represents a body with the maximum
compactness and, at the same time, it is the most conservative option
for such an ultra-compact object (lower, axial symmetry of the disc
radiation field limits the possibility of exploring the problem in an
analytical way).

Our paper takes general relativity effects into account, including
the effect of higher-order images if they arise. Although the
signal is usually weak in these images, favourable  geometrical
arrangements are possible and, even more importantly, photons of the
higher-order image experience a characteristic delay with  respect to
photons following a direct course. This delay (examined in detail by
\citealt{boz04}; \citealt{cad05}) could help revealing the presence of
strong gravitational field in the system.

Our model provides a useful test bed for astrophysically more realistic
schemes. In the next section we formulate the model and describe
calculations. Then we show comparisons with previous results of other
authors. We build our discussion on the approach of Beloborodov
(\citeyear{bel98}, for polarization) and Abramowicz et al.\
(\citeyear{abr90}, for the motion in combined gravitational and
radiation  fields in general relativity). In these papers the individual
components of the whole picture were treated separately, while we
connect them together in a consistent scheme. A reader interested only
in the main results on polarization of higher-order images from light
scattered on a moving cloudlet can proceed directly to
section~\ref{sec:cloudlet}.

As a final point of the introduction, it is worth noticing that
\citet{ghi04} propose a model of aborted jets in which colliding clouds
and shells occur very near a black hole and are embedded in strong
radiation field. According to their scheme, most of energy dissipation
should take place on the symmetry axis of an accretion disc. This would
be another suitable geometry, in which a fraction of light is boosted in
the direction to the photon circular orbit and eventually redirected to
the observer. One may fear that the Thomson scattering approximation is
not adequate to describe a turbulent medium in which electrons become
very hot \citep{pou94}, however, the accuracy should be still sufficient
for energies of several keV, at which planned polarimeters are supposed
to operate. Also the seed photons have different distribution when they
originate from an accretion disc, but we examined the variation of the
polarization that an observer can expect from this kind of a system
\citep{hor05}, and our calculations confirm that the expected polarization
has magnitude smaller but roughly similar to values predicted by a
simple model adopted here.

%-------------------------------------------------------------------------------
%  Polarized light from inverse Compton
%-------------------------------------------------------------------------------

\section{The model and calculations}
\subsection{The set up of the model and reference frames}
We assume fully ionized optically thin medium distributed outside the
source of seed photons. These primary photons follow null geodesics
until they are intercepted by an electron, which itself is moving under
the mutual competition between gravity and radiation (we do not consider
the effect of magnetic fields on the particle motion and radiation in
this paper). Hence, we adopt the approximation of single scattering and
we assume Schwarzschild metric for the gravitational field of a compact
body. We consider frequency-integrated quantities. Polarization vector
of scattered light is propagated  parallelly through gravitational field
to a distant observer and, as consequence, the polarization magnitude
$\Pi(\fvec{r},\fvec{n})$ and the redshifted intensity
$\tilde{I}\equiv\left(1+\mathcal{Z}\right)^{-4}I(\fvec{r},\fvec{n})$
(expressed here in terms of the redshift $\mathcal{Z}$) are invariant.

Polarization is described in terms of Stokes parameters $I$, $Q$, $U$
and $V$ \citep{cha60,ryb79}: $I$ has a meaning of intensity along a
light ray, $Q$ and $U$ characterize the linear polarization in two
orthogonal directions (say $\fvec{e}_X$ and $\fvec{e}_Y$) in the plane
perpendicular to the ray, and $V$ is the circularity parameter.  The
polarization angle is defined by $\tan\,2\chi=U/Q$. It gives the
orientation of major axis of the polarization ellipse with respect to
$\fvec{e}_X$. One can form a polarization basis in the local space by
supplementing $\fvec{e}_X$ and $\fvec{e}_Y$ with another unit vector,
$\fvec{e}_Z$, pointed in the direction of the light ray.   Three
parameters are necessary to describe a monochromatic beam, for which the
condition $I^2=Q^2+U^2+V^2$ holds. In case of partially polarized light
the whole set of four parameters is generally required. It is then
customary to define the degree of elliptical polarization,
$\Pi\,\equiv\,{I}^{-1}\,(Q^2+U^2+V^2)^{1/2}$, which satisfies
$0\leq\Pi\leq1$. 

\begin{figure*}
\begin{center}
\hfill~
\includegraphics[width=0.45\textwidth]{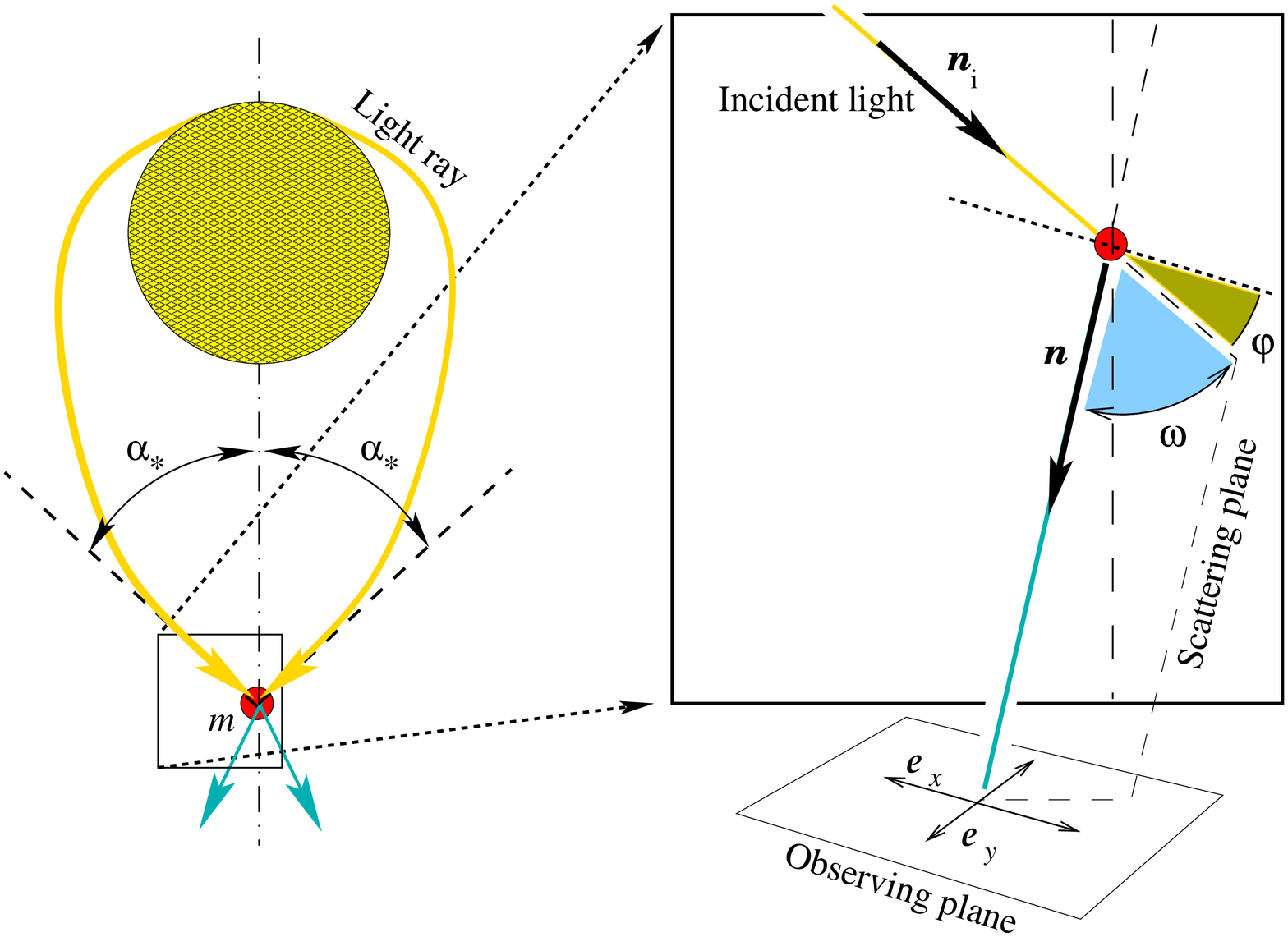}
\hfill~
\includegraphics[width=0.25\textwidth]{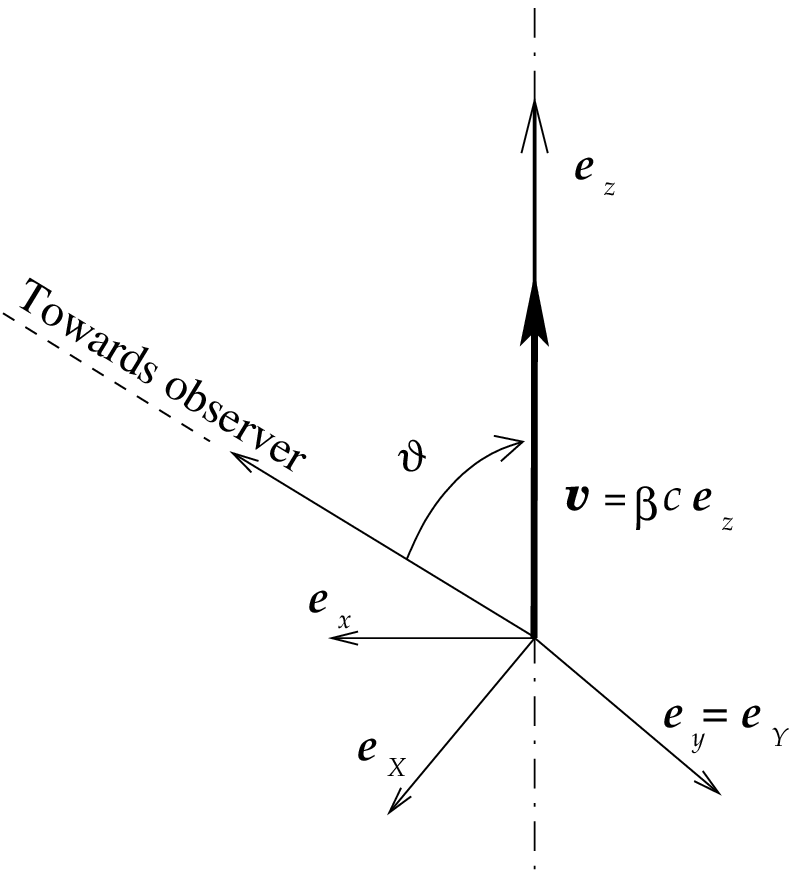}
\hfill~
\end{center}
\caption{Geometry of the problem and the definition of angles.}
\label{fig1}
\end{figure*}

A four-dimensional tetrad can be constructed by extending the three base
vectors and supplementing them with a purely timelike four-vector. A
suitable choice of the tetrad is described below. Let us consider a
simple case when the incident radiation field is axially symmetric in
the laboratory frame, LF, with the basis
$(\fvec{e}_t,\fvec{e}_x,\fvec{e}_y,\fvec{e}_z)$.\footnote{Hereafter we
understand three-vectors as spatial projections of their corresponding
four-vectors (and we do not introduce special notation for them;  there
is no danger of confusion). We stick to the conventional formalism of
Stokes parameters, but we remark that it can be recast by employing a
covariant definition of the polarization tensor, components of which are
assembled using suitable combinations of Stokes parameters
\citep{bor64}. It has been argued \citep{por05} that the latter approach
may be found more elegant and useful for discussing the radiation
transfer of polarized light through the medium in general relativity.}
Indices  of four-vectors with respect to a local-frame basis are
manipulated  by the flat-space metric, $\mathrm{diag}(-1,1,1,1)$. 

We orient $\fvec{e}_z$ along the symmetry axis; other two
spatial vectors, $\fvec{e}_x$ and $\fvec{e}_y$, lie in a plane
perpendicular to $\fvec{e}_z$. Further, we assume that 
scattering electrons are streaming along the symmetry axis with
four-velocity $\fvec{u}=u^t\fvec{e}_t+u^z\fvec{e}_z$, where 
$u^t=c\gamma$ and $u^z=c\gamma\beta$ ($\gamma$ is Lorentz factor,
$\beta$ is velocity in LF divided by speed of light). Later on we carry
out a Lorentz boost to the co-moving frame (CF) of the scatterer,
$(\bar{\fvec{e}}_t,\bar{\fvec{e}}_x,\bar{\fvec{e}}_y,\bar{\fvec{e}}_z)$,
which is equipped with timelike four-vector $\bar{\fvec{e}}_t=\fvec{u}$
and three space-like four-vectors $\bar{\fvec{e}}_x=\fvec{e}_x$,
$\bar{\fvec{e}}_y=\fvec{e}_y$. Spatial part of $\bar{\fvec{e}}_z$
is oriented in the direction of relative velocity of both frames. 

Each incident photon of the ambient unpolarized radiation gets
highly polarized when scattered by a relativistically moving electron.
The total polarization is eventually obtained by integrating over
incident directions  and the distribution of scattering electrons. 
In order to describe propagation of scattered photons, we denote
four-vectors $\fvec{n}\,\equiv\,\fvec{p}/p^t$ (with respect to LF) and 
$\bar{\fvec{n}}\,\equiv\,\fvec{p}/\bar{p}^t$ (with respect to CF), where
$\fvec{p}$ is the photon four-momentum (a null four-vector). Due to the
axial symmetry we can assume $n^y=\bar{n}^y=0$.

In addition to the above-defined reference frames LF and CF, we
introduce  two `polarization' frames: the lab polarization frame (LPF)
with basis $(\fvec{e}_t,\fvec{e}_X,\fvec{e}_Y,\fvec{e}_Z)$, and the
co-moving polarization frame (CPF) with the basis 
$(\bar{\fvec{e}}_t,\bar{\fvec{e}}_X,\bar{\fvec{e}}_Y,\bar{\fvec{e}}_Z)$. 
LPF is defined in such a way that $\fvec{e}_Z$ is the three-space
projection of the propagation four-vector $\fvec{n}$, $\fvec{e}_X$ lies
in the $(\fvec{e}_x,\fvec{e}_z)$-plane, and $\fvec{e}_Y$ is identical
with the LF tetrad vector $\fvec{e}_y$. CPF is defined analogically and
indicated by bars over variables. Our definition of the reference frames
is apparent from figure~\ref{fig1}.  

\subsection{Stokes parameters in terms of the incident radiation stress-energy tensor}
We start by calculating the polarization of the scattered radiation in 
CPF. Conceptually the model of local polarization is equivalent to the
one employed by \citet{bel98}. The incident radiation is unpolarized
with  intensity $\bar{I}_\mathrm{i}$. It can be imagined as a
superposition  of two parallel beams of identical intensities,
$\bar{I}_\mathrm{i}^{(1)}=\bar{I}_\mathrm{i}^{(2)}
=\bar{I}_\mathrm{i}/2$, propagating along $\bar{\fvec{n}}_\mathrm{i}$
four-vector. The two beams are completely linearly polarized in mutually
perpendicular directions and the scattered radiation is a mixture
of both components. In the adopted choice of reference frames (see the
figure~\ref{fig1}), the spatial projection of the propagation vector
$\bar{\fvec{n}}$ is identical with the spatial projection of
$\bar{\fvec{e}}_Z$. Unequal contributions $\bar{I}^{(1)}$ and
$\bar{I}^{(2)}$ to the total intensity $\bar{I}$ result in net linear
polarization of the scattered beam. According to this, $\bar{I}$,
$\bar{Q}$ and  $\bar{U}$ are non-zero, whereas the circularity parameter
$\bar{V}$ vanishes.

Assuming that each scattered photon experiences one scattering
event in an optically thin medium ($\tau\ll1$), non-zero contributions
to Stokes parameters are \citep{cha60}
\begin{eqnarray}
\delta \bar{I} &=& A \bar{I}_\mathrm{i} \left(1 + \cos^2\omega\right), \\
\delta \bar{Q} &=& -A \bar{I}_\mathrm{i} \,\cos 2\varphi \,\sin^2\omega, \\
\delta \bar{U} &=& -A \bar{I}_\mathrm{i} \,\sin 2\varphi \,\sin^2\omega,
\end{eqnarray}
where $\omega$ is the scattering angle between $\fvec{n}_\mathrm{i}$
and $\fvec{n}$, $A\,\equiv\,3\tau/(16\pi)$. The scattering  takes place in
the plane that forms an angle $\varphi$ with $\bar{x}$-axis. Angles
$\varphi$ and $\omega$ can be expressed using direction cosines,
which are defined here as  spatial components of the propagation
four-vector $\bar{n}_\mathrm{i}$ of the incident beam, i.e.\
$\bar{n}_\mathrm{i}^X=\cos\varphi\,\sin\omega$, 
$\bar{n}_\mathrm{i}^Y=\sin\varphi\,\sin\omega$ and 
$\bar{n}_\mathrm{i}^Z=\cos\omega$. We obtain
\begin{eqnarray}
\delta \bar{I} &=& A \left(1 + \bar{n}_\mathrm{i}^Z \bar{n}_\mathrm{i}^Z\right) 
\bar{I}_\mathrm{i}, 
\label{eq:I1} \\
\delta \bar{Q} &=& A \left(\bar{n}_\mathrm{i}^Y \bar{n}_\mathrm{i}^Y - 
\bar{n}_\mathrm{i}^X \bar{n}_\mathrm{i}^X\right) \bar{I}_\mathrm{i}, 
\label{eq:Q1} \\
\delta \bar{U} &=& -2A\, \bar{n}_\mathrm{i}^X \bar{n}_\mathrm{i}^Y 
\bar{I}_\mathrm{i}. 
\label{eq:U1}
\end{eqnarray}
This form is useful, as it allows integrating conveniently the
partial contributions over incident directions to obtain
\begin{eqnarray}
\bar{I} &\!\!\!=\!\!\!&
  Ac \left(\bar{T}^{tt} + \bar{T}^{ZZ}\right), 
  \label{eq:scattI} \\
\bar{Q} &\!\!\!=\!\!\!&
  Ac \left(\bar{T}^{YY} - \bar{T}^{XX}\right), 
  \label{eq:scattQ} \\
\bar{U} &\!\!\!=\!\!\!&
  -2Ac \bar{T}^{XY},
  \label{eq:scattU}
\end{eqnarray}
for the total Stokes parameters of scattered light. We denoted
\begin{equation}
\bar{T}^{\mu\nu}\equiv
\frac{1}{c}\int_{4\pi} \bar{n}_\mathrm{i}^\mu \bar{n}_\mathrm{i}^\nu 
\bar{I}_\mathrm{i}(\bar{\fvec{n}}_\mathrm{i})\,\mathrm{d}\Omega
\end{equation}
the stress-energy tensor of the incident radiation field.

\begin{figure*}
\includegraphics[width=0.45\textwidth]{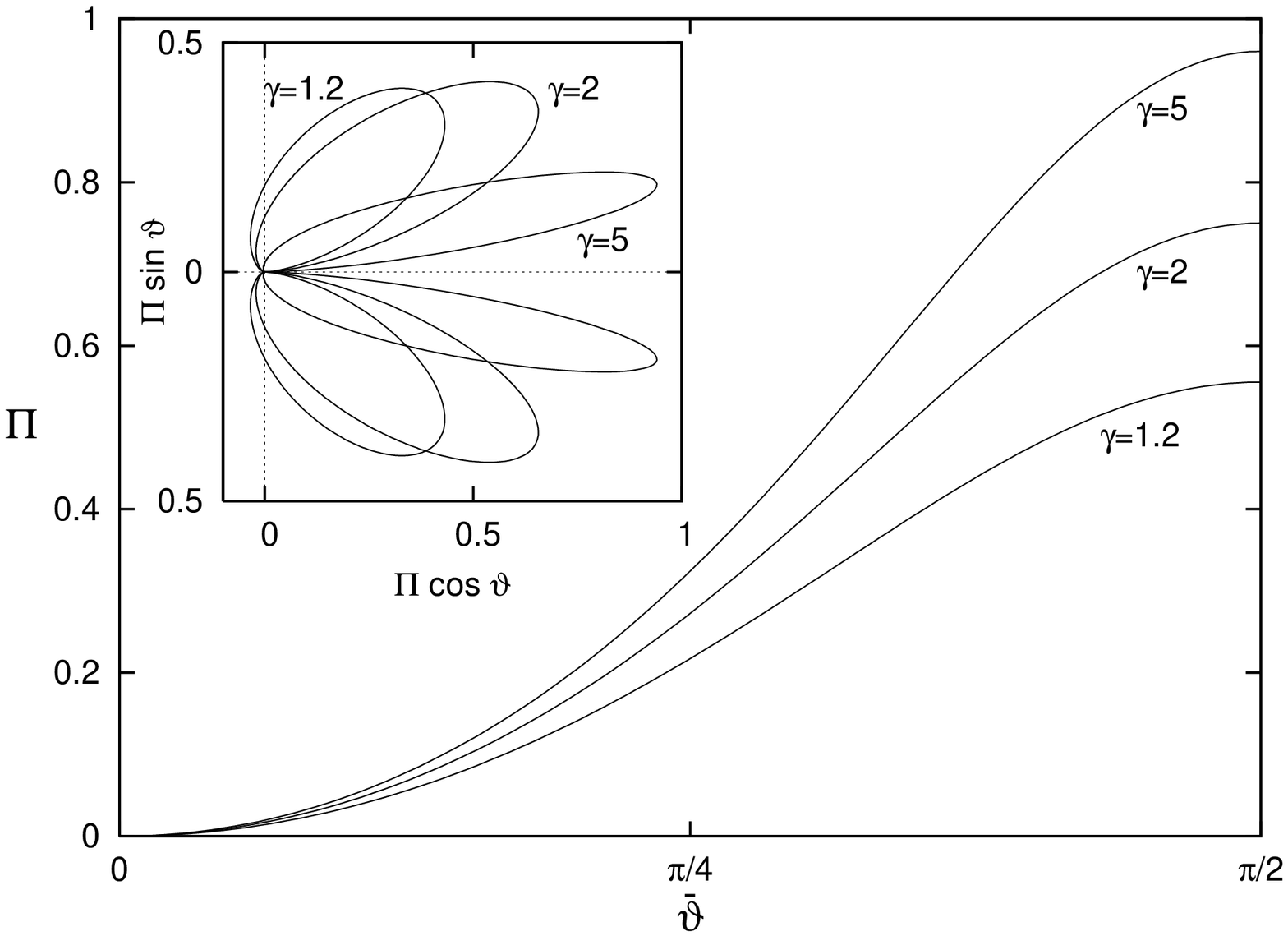}
\hfill
\includegraphics[width=0.54\textwidth]{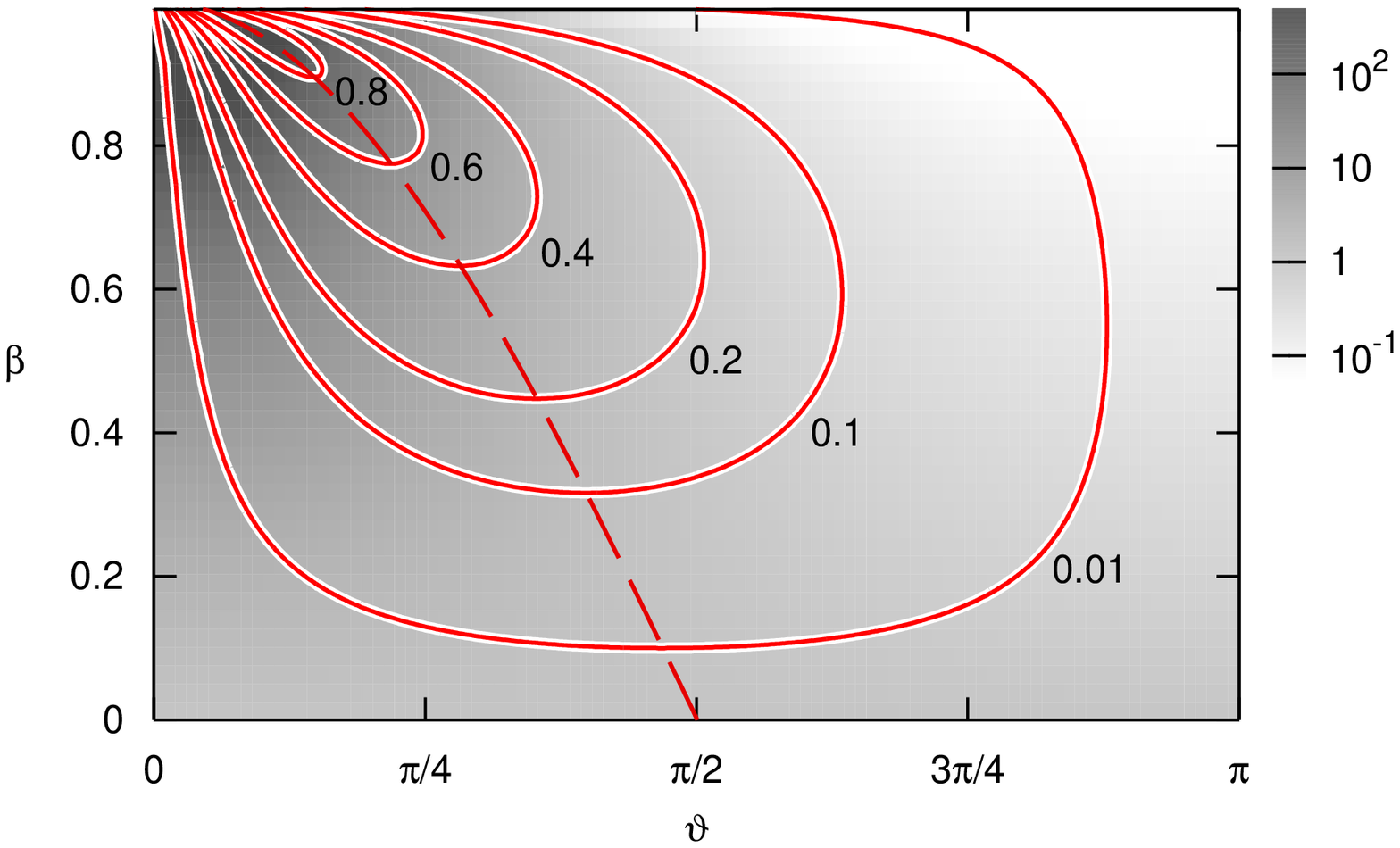}
\caption{Left: the magnitude of transversal polarization 
$\Pi(\bar{\vartheta};\gamma)$ due to up-scattering by a relativistic
electron as a function of observing angle in the local co-moving frame.
The case of locally isotropic ambient radiation field is shown for three
different values of Lorentz factor $\gamma$. In the inset the emission
diagram shows the corresponding lab-frame polarization. The lobes become
gradually flattened toward the front direction of motion as $\gamma$
increases. One can see from eq.~(\ref{eq:barI}) that the graph of
$\Pi(\bar{\vartheta})$ is symmetrical with respect to
$\bar{\vartheta}=\pi/2$. This corresponds to a well-known fact that
polarization is maximum for the radiation scattered perpendicularly to
axis of symmetry in the CF. Right: Contours of $\Pi(\vartheta,\beta)$
are shown superposed on the density plot of $I(\vartheta,\beta)$. Levels
of shading give the intensity (in arbitrary units) and illustrate the
progressive beaming towards $\vartheta=0$ direction in the
ultra-relativistic limit (in the lab frame). On the other hand, given a
value of $\beta$, the polarization degree $\Pi(\vartheta;\beta)$ as
function of $\vartheta$ reaches maximum at a non-zero angle, always off
axis (dashed line).}
\label{fig:iso}
\end{figure*}

We remind that the incident radiation was assumed axially symmetric in
the CF, therefore the only non-zero components in this frame are
$\bar{T}^{tt}$, $\bar{T}^{tz}$, $\bar{T}^{zz}$, $\bar{T}^{xx}$, and
$\bar{T}^{yy}$. These are further constrained by symmetry: 
$\bar{T}^{xx}=\bar{T}^{yy}=(\bar{T}^{tt}-\bar{T}^{zz})/2$. A relation to
CPF components can be obtained by rotation about $\bar{y}$-axis by angle
$\bar{\vartheta}$. Using equations (\ref{eq:scattI})--(\ref{eq:scattU})
we find
\begin{eqnarray}
 \bar{I} &\!\!\!=\!\!\!&
 \textstyle{\frac{1}{2}} Ac\big[\left(3\bar{T}^{tt}-\bar{T}^{zz}\right)-
 \left(\bar{T}^{tt}-3\bar{T}^{zz}\right)\cos^2\bar{\vartheta}\,\big], 
 \label{eq:barI}
 \\
 \bar{Q} &\!\!\!=\!\!\!& \textstyle{\frac{1}{2}}
 Ac\left(\bar{T}^{tt}-3\bar{T}^{zz}\right)\sin^2\bar{\vartheta}.
\end{eqnarray}
The Stokes parameter $\bar{U}$ vanishes due to axial symmetry.
The scattered radiation is partially polarized either in the
$(\bar{x},\bar{z})$-plane, or perpendicularly to it. The former  and the
latter case will be referred as longitudinal and transverse
polarization, respectively. Later on, in
Sec.~\ref{sec:relstar}, we will demonstrate that this change
can occur also with a group of cold electrons, 
whose bulk motion is determined by the radiation and gravitational fields 
of a star and recorded in the lab frame.

The degree of polarization is calculated directly from definition:
\begin{equation}
\Pi(\bar{\vartheta})=\frac{|\bar{Q}|}{\bar{I}} 
  =\frac{|\Pi_\mathrm{m}|\sin^2\bar{\vartheta}}
  {1 - \Pi_\mathrm{m}\cos^2\bar{\vartheta}}\,,
\label{eq:polCF}
\end{equation}
where 
\begin{equation}
\Pi_\mathrm{m}\equiv\frac{\bar{T}^{tt}-3\bar{T}^{zz}}
  {3\bar{T}^{tt}-\bar{T}^{zz}}.
\end{equation}
This result is equivalent to eqs.~(4)--(5) in \citet{bel98}, who applied
the model of Thomson scattering to winds outflowing from a
plane-parallel disc  slab. Beloborodov also calculated the polarization
of scattered radiation and he found that its sign depends on the wind 
velocity. In our notation the change is captured in $\Pi_\mathrm{m}$,
which acquires values in the range $-1\leq\Pi_\mathrm{m}\leq1$. Its
meaning is evident from eq.~(\ref{eq:polCF}): the absolute value
$|\Pi_\mathrm{m}|$ is the maximum degree of  polarization of the
scattered light and the sign of $\Pi_\mathrm{m}$ determines the sign of
$\bar{Q}$-parameter.

In order to determine the polarization magnitude as seen by an 
observer in LF we carry out the Lorentz boost (e.g.,
\citealt{coc72}; \citealt{ryb79}). The angle of observation 
$\bar{\vartheta}$ is transformed according to: 
\begin{equation}
\sin\bar{\vartheta}=\mathcal{D}\;\sin\vartheta,\quad
\cos\bar{\vartheta}=\gamma\mathcal{D}\;(\cos\vartheta-\beta), 
\end{equation}
where  $\mathcal{D}\,\equiv\,\gamma^{-1}(1-\beta\cos\vartheta)^{-1}$ is the Doppler
factor. Stokes parameters are transformed in the same
manner as the radiation intensity and the boost retains the
four-vector $\fvec{e}_y$ unchanged:
$I=\mathcal{D}^4\bar{I}$ and $Q=\mathcal{D}^4\bar{Q}$.
It follows that the polarization magnitude $|\Pi_\mathrm{m}|$ is Lorentz
invariant. By transforming all relevant quantities to LF, we obtain 
Stokes parameters of scattered radiation,
\begin{eqnarray}
Q&=&\textstyle{\frac{1}{2}}Ac\,\mathcal{D}^6\gamma^2
 \big[(1-3\beta^2)T^{tt} 
\nonumber\\
  && +4\beta T^{tz} - (3-\beta^2)T^{zz}\big]\sin^2\vartheta, \\
  I&=&Ac\mathcal{D}^4\gamma^2\,\big[(1+\beta^2)\left(T^{tt}+T^{zz}\right) - 
  4\beta T^{tz}\big] + Q.
\label{eq:isc}
\end{eqnarray}

%-------------------------------------------------------------------------------
%  Critical velocities
%-------------------------------------------------------------------------------

\subsection{Critical velocities}
The aim of this subsection is to connect, in a self-consistent manner, the
properties of particle motion through the ambient radiation field with
Stokes parameters of scattered light. In order to prepare for this
discussion it is useful to introduce two critical velocities of the
particle motion.

Firstly, of particular interest is the velocity at which the
polarization of scattered radiation vanishes \citep{bel98}. 
The condition for velocity follows from the requirement
\begin{equation}
\bar{T}^{tt} - 3\bar{T}^{zz} = 0.
\label{eq:pi00}
\end{equation}
Performing the Lorentz boost to LF we obtain
\begin{eqnarray}
\left(1-3\beta^2\right)T^{tt} + 4\beta T^{tz}
 + \left(\beta^2-3\right)T^{zz} = 0.
\end{eqnarray}
This is a quadratic equation for $\beta$, which has two roots,
\begin{eqnarray}
\beta_{1,2} = a \pm \sqrt{a^2 + b}\,,
\end{eqnarray}
where
\begin{eqnarray}
a\equiv\frac{2T^{tz}}{3T^{tt}-T^{zz}}\,,
\quad
b\equiv\frac{T^{tt}-3T^{zz}}{3T^{tt}-T^{zz}}\,.
\end{eqnarray}
Clearly, eq.~(\ref{eq:pi00}) can be satisfied independently of the
direction of observation. For $\beta\rightarrow\beta_{1,2}$ the
polarization changes from longitudinal to transversal.

Secondly, we introduce the saturation velocity $\beta_0$ \citep{sik81}. 
As was shown by various authors under different approximations about the
particle cross-section and the form of gravitational field (see e.g.\
\citealt{abr90}; \citealt{vok91}; \citealt{mel89}; \citealt{fuk99};
\citealt{kea01}), the saturation velocity plays an important role in the
dynamics of relativistic jets: particles moving at velocity smaller/greater 
than the saturation velocity gain/lose their momentum at the expense of
the radiation field. In absence of other acceleration mechanisms and
neglecting inertia of particles, the effect of radiation pressure
eventually leads to $\beta\rightarrow\beta_0$ as terminal speed of the
particle motion. 

The saturation velocity is determined by the requirement of the
vanishing radiation flux in CF, i.e.
\begin{equation} 
  \bar{T}^{tz}=0. 
\end{equation}  
This gives another quadratic equation,
\begin{equation}
\left(1+\beta^2\right)T^{tz}-\beta\left(T^{tt}+T^{zz}\right)=0,
\end{equation}
with the solution 
\begin{equation}
\beta_0=\frac{1-\sqrt{1-\sigma^2}}{\sigma}\,, 
\label{eq:beta0}
\end{equation}
where $\sigma\,\equiv\,2T^{tz}/(T^{tt}+T^{tz})$.
We ignore the second solution, as it has no physical meaning.

%-------------------------------------------------------------------------------
%  Example
%-------------------------------------------------------------------------------

%\subsection{An example}
\label{sec:example}
As an example let us assume the incident radiation field to be strictly 
isotropic in the laboratory frame, i.e.
\begin{equation}
T^{\alpha\beta} = \mathrm{diag}\left(\mathcal{E},
 \textstyle{\frac{1}{3}}\mathcal{E},
 \textstyle{\frac{1}{3}}\mathcal{E},
 \textstyle{\frac{1}{3}}\mathcal{E}\right)
\label{eq:iso1}
\end{equation}
with $\mathcal{E}{\,\equiv\,}T^{tt}$ being energy-density of radiation. 
Evaluating the stress-energy tensor in CF we find
$\Pi_\mathrm{m}=-\beta^2$. Substituting into the
equation~(\ref{eq:polCF}) we obtain polarization degree
\begin{equation}
\Pi(\bar{\vartheta},\beta) =
 \frac{\beta^2\sin^2\bar{\vartheta}}{1+\beta^2\cos^2\bar{\vartheta}}\;.
\label{eq:pi1}
\end{equation}
Lorentz transformation to LF gives
\begin{equation}
\Pi(\vartheta,\beta) = \frac{\beta^2\sin^2\vartheta}{\left(2\gamma^2-1\right)
 \left(1-\beta\cos\vartheta\right)^2 - \beta^2\sin^2\vartheta}\;.
\label{eq:pi2}
\end{equation}
Since $\Pi_\mathrm{m}\leq0$, the scattered radiation is  polarized
transversely. The critical velocities are  $\beta_0=\beta_1=\beta_2=0$
in this case.

Figure~\ref{fig:iso} shows the dependence of $\Pi$ on observing angle
according to equation~(\ref{eq:pi1}). It is worth noticing that we
explore the frequency-integrated model, because this assumption is
adequate for the purpose of clarification of the role gravitational
lensing (discussed in the next section). The same dependency of Stokes
parameters on the scatterer velocity and observer viewing angle is
obtained in frequency-dependent calculation with the spectral index of
the incident radiation equal to $-1$ (the case adopted originally by
\citealt{beg87}). Our results are consistent with
\citet{laz04} provided that appropriate averaging over energy is
adopted. It can be seen (in the left panel of Fig.~\ref{fig:iso}) that
the resulting curves closely resemble the numerical result of Lazzati et
al.\ (\citeyear{laz04}; cp.\ their Fig.~1). In particular, the
curves are identical for $\gamma\gg1$ and they approach the
ultra-relativistic limit
$\Pi=(1-\cos^2\bar{\vartheta})/(1+\cos^2\bar{\vartheta})$ (originally
examined again by \citealt{beg87}, and invoked more recently e.g.\ by
\citealt{sha95}). This limit corresponds to the case of a head-on
collision, when all photons are impinging at incident angles
$\bar{\vartheta}_\mathrm{i}\rightarrow\pi$ because of aberration in CF.

For moderate Lorentz factors there is some difference between our
profile of $\Pi(\bar{\vartheta})$ and the corresponding numerical values
plotted in \citet{laz04}. For example, checking the $\gamma=2$ curve, we
notice that relative difference amounts to roughly $13$\%. This
apparent discrepancy is explained by realizing that our
eq.~(\ref{eq:pi1}) has been derived in terms of bolometric intensities,
whereas Lazzati et al.\ employ specific (frequency-dependent)
quantities. By integrating their Stokes parameters over frequency we
recover precisely the value predicted by eq.~(\ref{eq:pi1}). See
\citet{hor05} for detailed comparisons.

\begin{figure*}
\includegraphics[width=0.49\textwidth]{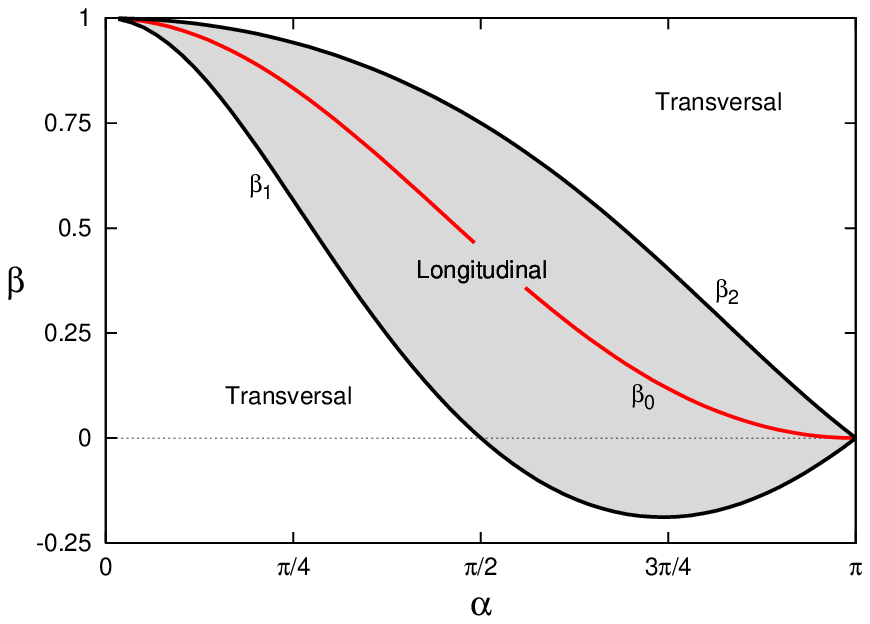}
\hfill
\includegraphics[width=0.49\textwidth]{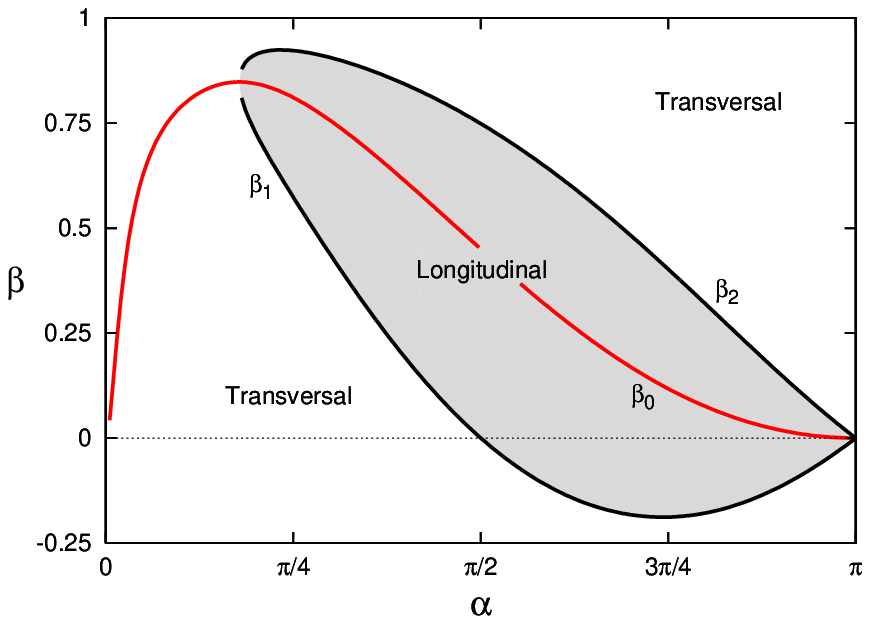}
\caption{Left: 
The case of incident radiation originating from an isotropic source of 
angular radius $\alpha$; see eqs.~(\ref{eq:estar})--(\ref{eq:pstar}).
Two branches of critical velocity are shown, $\beta_1(\alpha)$ and
$\beta_2(\alpha)$, at which the total polarization of scattered light
vanishes independent of observing direction. The saturation curve
$\beta_0(\alpha)$ is also plotted assuming that the radiation drag
dominates the particle dynamics. Right: the same as on the left but for
a mixture of two components of the incident radiation, i.e.\ the ambient
isotropic ($\alpha=\pi$) source plus stellar (non-isotropic)
contribution according  to eq.~(\ref{eq:lam1}) with
$\lambda_\mathrm{i}=0.001$. In both panels, the regions of longitudinal
and transversal polarization are distinguished by shading.}
\label{fig4}
\end{figure*}

%-------------------------------------------------------------------------------
%  Polarization of light near a compact star
%-------------------------------------------------------------------------------

\section{Polarization of light scattered near a compact star}
\label{sec:relstar}
\subsection{The gravitational and radiation fields}
The gravitational field of a spherically symmetric star is described by 
Schwarzschild metric \citep{cha92},
\begin{equation}
{\rm d}s^2 = -c^2\xi\, {\rm d}t^2 + \xi^{-1}{\rm d}r^2 + r^2\,{\rm d}\Omega^2,
\label{eq:ds}
\end{equation}
where ${\mathrm{d}}\Omega$ is the angular part of a spherically
symmetric line element, $\xi(r)\,\equiv\,1-\rs/r$ is
the redshift function in terms of Schwarzschild radius,
$\rs\,\equiv\,{2GM}c^{-2}\dot{=}2.95\times10^5(M/M_{\sun})$~cm, and $M$ is
mass of the star. 
Four-vectors and four-tensors will be
expressed with respect to  a local orthonormal tetrad,
$(\fvec{e}^{(t)},\fvec{e}^{(r)},\fvec{e}^{(\theta)},\fvec{e}^{(\phi)})$,
with non-vanishing components $e^{(t)}_t=c\xi^{1/2}$,
$e^{(r)}_r=\xi^{-1/2}$,  $e^{(\theta)}_\theta=r$ and
$e^{(\phi)}_\phi=r\sin\theta$. Tetrad components of four-vectors  are
denoted by bracketed indices and are raised and lowered using the
Minkowski metric. 

Primary photons are emitted from a star 
and form the ambient radiation field acting on the particle. The star of
radius $R_\star$ and compactness $R_\star/\rs$ appears
to a static observer, located at radius $r$,
as a bright disc of angular radius $\alpha_\star=\alpha(r)$,
\begin{equation}
\sin\alpha(r)=\frac{\tilde{R}}{r}\;\frac{\xi(r)^{1/2}}{\xi(\tilde{R})^{1/2}},
\label{eq:angle}
\end{equation}
where $\tilde{R}\,\equiv\,\max\{\textstyle{\frac{3}{2}}\rs,R_\star\}$. 
Because of light bending the solid angle subtended by a compact star on
the sky is larger than the Euclidean (flat space) estimate.
Formula~(\ref{eq:angle}) was originally discussed by  \citet{syn67} and
the manifestation of the gravitational self-lense effect was examined
by \citet{win73}.

In case of very high compactness (when
$R_\star<\frac{3}{2}\rs$) the rim of the image is formed by 
photons encircling the perimeter of the star more than once. In 
spite of complicated trajectories of photons, to the observer 
the surface appears radiating with intensity (we neglect limb darkening 
for simplicity)
\begin{equation}
I(r) = \frac{\xi(R_\star)^2}{\xi(r)^2}\;I_\star(R_\star).
\label{eq:I}
\end{equation}
Let us take the previous example from eq.~(\ref{eq:iso1}), but assume
that the source of primary photons occupies only a fraction of the
local sky of the scattering particle. Gradual dilution of the source
radiation with distance is described by function $\alpha(r)$. The limit
of $\alpha\rightarrow\pi$ corresponds to strictly isotropic radiation
arriving from all directions, whereas for $\alpha\rightarrow0$ we
obtain the case of a point-like source. We thus recognize the results of
subsection~\ref{sec:example}.

The stress-energy tensor of the stellar radiation field has three
independent components, namely, the energy density, the energy flux, and
the radial stress. These are given, respectively, by
\begin{eqnarray}
\mathcal{E}_\star &\equiv& T_\star^{(t)(t)} = 
 \frac{2\pi}{c} I \left(1-\cos\alpha\right), \label{eq:estar} \\
\mathcal{F}_\star &\equiv& cT_\star^{(t)(r)} = \pi I \sin^2\alpha, \\
\mathcal{P}_\star &\equiv& T_\star^{(r)(r)} = \frac{2\pi}{3c}
 I \left(1-\cos^3\alpha\right). \label{eq:pstar}
\end{eqnarray}
There are two other non-zero components,
$T_\star^{(\theta)(\theta)}=T_\star^{(\phi)(\phi)}$, which can be
computed from the condition $T_{\star \sigma}^\sigma=0$. Magnitude of 
the star radiation is characterized by total luminosity,
$L_\star = 4\pi R_\star^2 \mathcal{F}_\star(R_\star)$.

Finally we include another, isotropic component of the radiation field 
with intensity $I_\mathrm{iso}$ in addition to stellar light. The
corresponding stress-energy tensor is entirely determined by energy
density $\mathcal{E}_\mathrm{iso}=4{\pi}c^{-1}I_\mathrm{iso}$.
The stress-energy tensor of the total radiation field is the sum 
$T^{\alpha\beta}=T_\star^{\alpha\beta} + T_\mathrm{iso}^{\alpha\beta}$.
Combining the two contributions allows us to model different 
situations according to their relative magnitude and the motion 
of the scattering medium.

\begin{figure*}
\includegraphics[width=\textwidth]{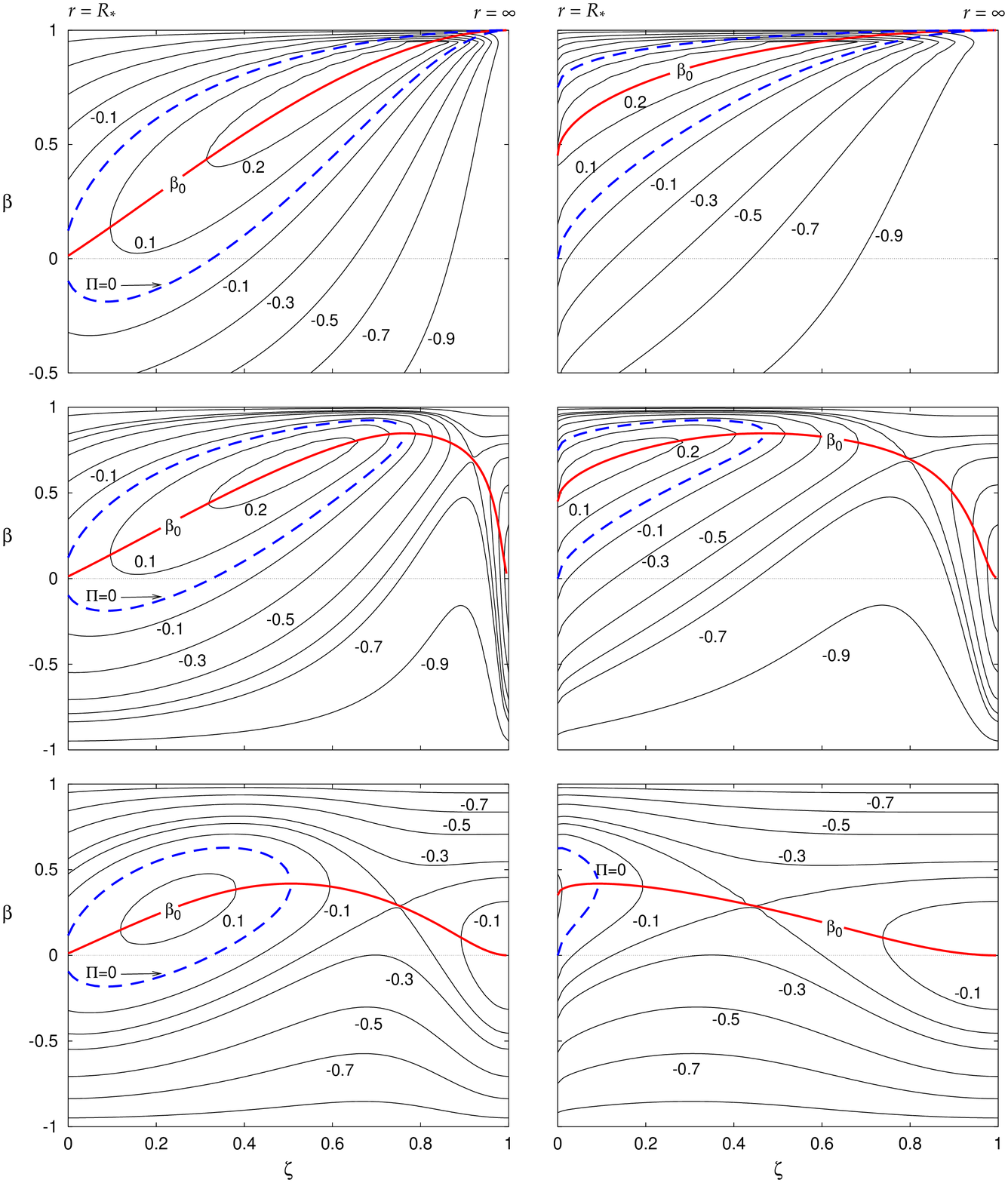}
\caption{Contours of extremal values of the polarization function
$\Pi_{\mathrm{m}}(\beta,\zeta)$ for photons Thomson-scattered on an
electron, moving with a given velocity $\beta$ through a mixture of
stellar and ambient diffuse light.  Each panel captures the whole range
of radii from $r=R_\star$ ($\zeta=0$) to $r\rightarrow\infty$
($\zeta=1$). The three rows correspond to progressively increasing
luminosity parameter (\ref{eq:lam1}):  (i)~$\lambda_{\mathrm{i}}=0$
(top); (ii)~$\lambda_{\mathrm{i}}=0.001$ (middle);  and
(iii)~$\lambda_{\mathrm{i}}=0.1$ (bottom). The left column is for a
highly compact star with $R_\star=1.01\rs$; the right column is for
$R_\star=10^3\rs$. Hence, the light-bending effects are significant on
the left and negligible on the right. The curve of zero polarization
$\Pi=0$ is plotted with a dashed line. Generally, if the star is
sufficiently compact then the curve of zero polarization becomes
double-valued; its two branches correspond to
$\beta\,\equiv\,\beta_{1,2}(\zeta)$ in the previous figure. A separatrix
is a particular contour that distinguishes regions of different topology
in this graph. The saturation curve $\beta_{\mathrm{}0}(\zeta)$ is also 
plotted.}
\label{fig5}
\end{figure*}

\subsection{Polarization of scattered light}
\label{sec:scattered}
We first assume velocity of the scatterer $\beta(r)$ and compute the
resulting polarization.  Figure~\ref{fig4} shows the effect of vanishing
and changing  polarization which occurs at particular value of $\beta$.
We compare two situations: the case of purely stellar component of the
primary irradiation, as described in the previous paragraph (in the left
panel), versus the case of a sum of the isotropic component and
radiation coming from the star surface (on the right). The latter
configuration represents an anisotropic irradiation of an electron;
it can be parametrized by the mutual ratio of redshifted radiation
intensities received at a distant observer, i.e. 
\begin{equation}
\lambda_{\mathrm{i}}\equiv
\frac{\tilde{I}_\mathrm{iso}}{\tilde{I}_{\star}}.
\label{eq:lam1}
\end{equation}
This can be considered as a toy-model of inverse Compton up-scatter in
an illuminated jet where intensity of ambient light is not directly
connected with the intensity of the central source.
$\lambda_{\mathrm{i}}=\mathrm{const}$ is a free parameter of the model;
given a value, the degree of anisotropy depends on the distance from the
star in $\rs$. It is worth noticing that light-bending is taken into
account in this calculation automatically, including all higher-order
images encircling the star.

Polarization is non-zero provided that particle velocity is not equal to
$\beta_{1,2}(r)$ and, indeed, $\Pi$ can reach large values. This is
shown in figure~\ref{fig5}, where we plot the extremal value of function 
$\Pi_{\mathrm{m}}(\beta,\zeta)$ in the plane of particle velocity versus
distance. $|\Pi_{\mathrm{m}}(\beta,\zeta)|$ is equal to extremes of the
polarization degree measured along a suitably chosen observing angle
$\vartheta$. The curve of zero polarization is also plotted and we
notice that it is independent of $\vartheta$. In this figure the primary
unpolarized light was assumed to be a mixture of stellar and ambient
contributions (the latter component was assumed to be distributed
isotropically in the lab frame). The saturation curve $\beta_0(\zeta)$
is shown and it is worth noticing that, for some values of the model
parameters, $\beta_0(\zeta)$ crosses the contour of $\Pi=0$. Therefore,
a hypothetical particle moving or oscillating along the saturation curve
would exhibit polarization that swings its direction by right angle.

One can modify the previous example by considering a constant ratio
of energy density, i.e.\ by replacing $\lambda_{\mathrm{i}}$ with
another parameter,
\begin{equation}
\lambda_{\mathrm{e}}\equiv\frac{\mathcal{E}_\mathrm{iso}}{\mathcal{E}_{\star}(r)}.
\label{eq:lam2}
\end{equation}
This definition captures better the case when the ambient light originates
from scattering of the central component (perhaps on clumps being accreted
onto the star), so that both contributions are linked to each other and
their energy density decreases at identical rate with the distance. We
again constructed graphs of $\Pi_\mathrm{m}(\beta,\zeta)$ and found a
similar structure of contours at small radii as those shown in
Fig.~\ref{fig5}, including the double-valued function
$\beta_{1,2}(\zeta)$. However, the saturation velocity $\beta_0(\zeta)$
does not fall to zero at $r\rightarrow\infty$, instead, it generally
reaches substantially higher values. Moreover, the critical point (where
the separatrix curve self-crosses) is lost, as well as the whole
structure towards right of the critical point. 

Polarization of scattered light obviously depends on scatterer motion
and these can be calculated consistently. In the next section we finally
determine velocity $\beta(r)$ along the particle trajectories, for which
luminosity of the star and its compactness are parameters.

\subsection{Polarization along the particle trajectory}
\label{sec:dynamics}
Four-velocity $\fvec{u}$ of a particle can be found by solving the equation
of motion in the form \citep{abr90},
\begin{equation}
mu_\rho \fvec{\nabla}^\rho u^\alpha
=-\frac{\sigma_{_\mathrm{T}}}{c}\,h^{\alpha}_{\rho}\,T^{\rho\sigma}u_\sigma,
\label{eq:eom}
\end{equation}
where $m$ is the particle rest mass and
$h^\nu_\mu\,\equiv\,\delta^\nu_\mu+c^{-2}u^\nu u_\mu$ is a projection
tensor. Left-hand side of eq.~(\ref{eq:eom}) includes the effect of
gravity ($\fvec{\nabla}$ denotes covariant differentiation with respect
to curved spacetime geometry) and the right-hand side provides the
effect of radiation drag -- accelerating or decelerating particles with
respect to free-fall motion. 

\begin{figure*}
\includegraphics[width=\textwidth]{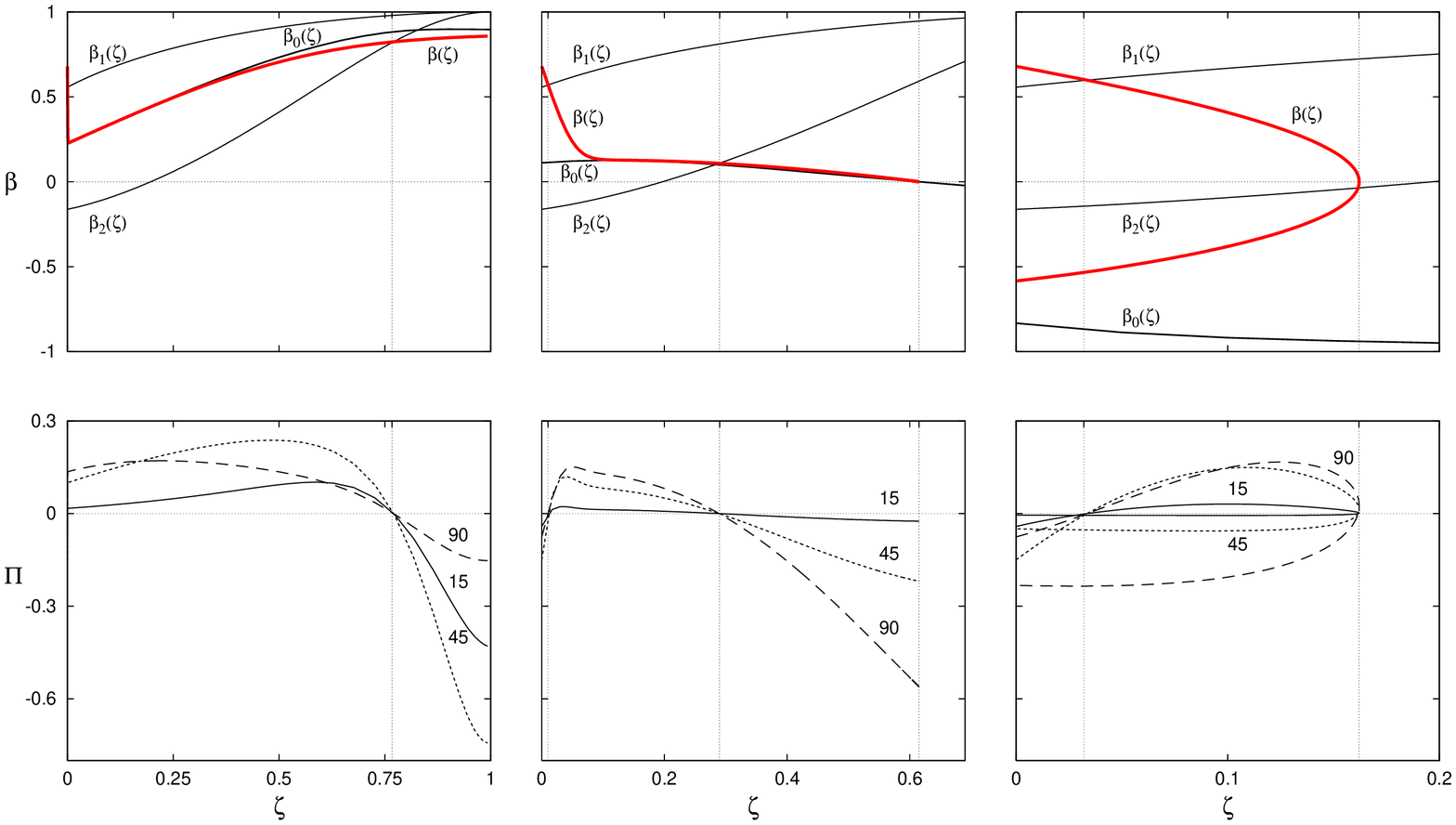}
\caption{Top row: particle velocity $\beta(\zeta)$ (thick curve) and
the three critical velocities $\beta_0$, $\beta_1$ and $\beta_2$ in the
combined radiation and gravitational fields. Each trajectory starts from
the star, $R_\star=1.205\rs$ ($\zeta=0$), and it proceeds
towards infinity ($\zeta=1$). Positive values of $\beta$  correspond to
an outflow, negative values are for an inflowing material. Three cases
are shown for different values of dimensionless luminosity: $\Gamma=100$
(left), $\Gamma=2$ (middle), and $\Gamma=0.1$ (right). Bottom row:
the polarization magnitude $\Pi(\zeta)$ along the particle trajectories
corresponding to the three above-given solutions. In each panel, the
curves are labelled with the observing angle $\vartheta$. All curves have a
common zero point where they cross each other. The sign of $\Pi$
distinguishes here the case of transversal polarization from the
longitudinal one.}
\label{fig:trajectories}
\end{figure*}

Non-zero components of four-velocity are $u^{(t)}=c\gamma$ and 
$u^{(r)}=c\gamma\beta$ in the local tetrad of a static
observer.\footnote{In this paper we consider purely radial motion. The
same formalism can be readily applied to more complicated motion of the
scatterer, although it will then hardly be possible to solve both the
motion and the resulting polarization analytically. Notice that the
case of clumps orbiting in the plane of a black hole accretion disc was
discussed by \citet{pin77}, \citet{con80} and \citet{bao97}. These
authors also pointed out that effects of general relativity could be
discovered by tracing time variable polarization.} 
Equation~(\ref{eq:eom}) takes the form
\begin{eqnarray}
  \dot{\beta}&=&
  \frac{1}{c\gamma^2}\left(\frac{\xi^{1/2}}{\gamma^2}\frac{F^{(r)}_\mathrm{rad}}{m} 
  - \frac{c^2\rs}{2r^2}\right),
  \label{eq:eom-beta} \\
  \dot{r}&=&c\xi\beta,
  \label{eq:eom-r}
\end{eqnarray}
where dot denotes derivative with respect to coordinate time $t$ and 
$F^{(r)}_\mathrm{rad}$ is the radial component of the radiation force
\begin{equation}
  F^{(r)}_\mathrm{rad} = \sigma_{_\mathrm{T}}\gamma^3
  \left[\left(1+\beta^2\right)T^{(t)(r)} - 
  \beta\left(T^{(t)(t)} + T^{(r)(r)}\right)\right].
  \label{eq:eom-F}
\end{equation}
The effect of radiation on the motion is expressed by the first term in
the parentheses on the right-hand side of eq.~(\ref{eq:eom-beta}),
whereas the other term can be considered as the contribution of gravity. 
Hence, the particle dynamics depends on relative strength of the
radiation  and gravitational fields. Because of redshift factor near a
compact star  these two influences do not obey the same simple Newtonian
law, and so a rich set of possible results emerges. These can be
parametrized by Eddington luminosity $L_\mathrm{E}$, which follows from
the condition of zero acceleration for matter hovering at
radius $r=R_{\star}$. The radiation force becomes
$F^{(r)}_\mathrm{rad}=(\sigma_{_\mathrm{T}}L_\star)/(4{\pi}R_{\star}^2c)$ 
and the equation (\ref{eq:eom-beta}) with $\dot{\beta}=0$ gives
\begin{equation}
L_\mathrm{E}=\frac{2\pi mc^3\rs}{\sigma_{_\mathrm{T}}\xi(R_\star)^{1/2}}.
\end{equation}
We note, that the acceleration depends on the radial distance from the
center as well as on particle velocity. The relative importance of
radiation and gravity is characterised by dimensionless factor
\begin{equation}
\Gamma \equiv \frac{L_\star}{L_\mathrm{E}}.
\end{equation}
The radiation term in acceleration is regulated by the interplay of
relativistic aberration and the Doppler boosting, which tend to
establish the saturation velocity. At this point further acceleration
vanishes, i.e.\ $\dot{\beta}_0(r)=0$. Considering only radiation from
the star and expressing the explicit form of the stress-energy tensor, 
eqs.~(\ref{eq:eom-beta})--(\ref{eq:eom-r}) reduce to eq.~(2.3) of 
\cite{abr90}. With gravitational attraction of the centre taken into
account, a possibility occurs of an equilibrium point
$\zeta_{\rm{}eq}\,\equiv\,\zeta(r=R_{\rm{}eq})$, where a particle can
reside. By setting $\beta=0$ and $\dot{\beta}=0$ in
eqs.~(\ref{eq:eom-beta})--(\ref{eq:eom-F}), one can find that the
equilibrium radius ranges from $R_{\rm{}eq}(\Gamma){\rightarrow}R_\star$
(i.e.\ $\zeta_{\rm{}eq}\rightarrow0$) for $\Gamma\rightarrow1$ up to
$R_{\rm{}eq}(\Gamma)\rightarrow\infty$ (i.e.\
$\zeta_{\rm{}eq}\rightarrow1$) for $\Gamma\rightarrow\sqrt{3}$.

Equations (\ref{eq:eom-beta})--(\ref{eq:eom-r}) allow for a finite set
of topologically different solutions. These can be classified into
different categories (\citealt{abr90}; see also \citealt{kea01})
according to the behaviour of saturation curves in
$(\beta,\zeta)$-plane. Notice that we already examined one of these
solutions, i.e.\ the saturation curve (\ref{eq:beta0}) for very high
luminosity of the star and negligible mass of the particle, i.e.,
$\Gamma\rightarrow\infty$.
The motion is then governed solely by radiation drag. We select this
condition because it is particularly relevant for the discussion of the
resulting polarization of scattered light. Its role can be
inferred also from Fig.~\ref{fig5}, where the $\beta_0$-curve for this
case  passes through the critical point of contour lines of
$\Pi(\beta,\zeta)$. The limit of $\Gamma\rightarrow\infty$ is an extreme
case. Different profile $\beta_0(\zeta)$ applies to moderate values of
the luminosity parameter, $\Gamma<\infty$, when particles do not
strictly maintain the saturation velocity because of inertial effects
acting on them. 

Different categories of the particle motion then provide a natural
framework also for the discussion of the resulting polarization. Three most
important cases are recorded in figure~\ref{fig:trajectories}. In this
example only the stellar radiation is taken into account, whereas the
component $I_\mathrm{iso}=0$ for simplicity. The cases shown here
correspond to the situation when (i)~the radiation field dominates over
gravity and the electron is therefore pushed away to an infinite radius (see
the left panel); (ii)~a moderate value of the luminosity allows the scattering
electrons to reach an equilibrium position at
$\zeta\,\equiv\,\zeta_{\rm{}eq}=0.62$ (middle);
(iii)~the luminosity is very small and the particles are almost
free-falling in the gravitational field (right). Particles
start from $\zeta=0$ and they quickly adhere
to the saturation curve $\beta_0(\zeta;\Gamma)$, provided that radiation 
is dynamically important, i.e.\ in cases (i) and (ii). This occurs
independently of initial velocity; then the motion follows a curve
adjacent to but slightly different from the saturation curve. On the
other hand, in the case (iii) the gravitation governs the motion; the
trajectory $\beta(\zeta)$ is only slightly asymmetric with respect to
$\beta=0$ line by the weak influence of radiation.

By coupling the equations of particle motion with the polarization
equations of sec.~\ref{sec:scattered} we obtain Stokes parameters of
scattered light along each particle trajectory. Bottom panels of
Fig.~\ref{fig:trajectories} show the resulting magnitude of
polarization. Notice how it crosses zero level at certain distance of
the scatterer from the stellar surface. At this point polarization
changes direction from transverse to longitudinal. Points of
intersection of curves $\beta_{1,2}(\zeta)$ with the particle motion
$\beta(\zeta)$ determine the radial location of the point of vanishing
polarization (indicated by dotted vertical lines in the plot).

We will now assume that a small cloudlet is formed by a group of
electrons. We can distinguish three cases, depending on the bulk
velocity of the cloudlet. These are discussed in subsections
\ref{sec:outward}--\ref{sec:inward} below. The predicted time dependence
offers a way to test the model.

\section{Polarization from a cloudlet}
\label{sec:cloudlet}

\subsection{The case of fast ejection ({\mbox{\protect\boldmath{$\gamma\gg1$}}})}
\label{sec:outward}
Let us denote $R_\mathrm{cl}\ll\rs$ radius of the cloud and 
$\psi{\sim}R_\mathrm{cl}/z$ its angular radius as seen from the
center. We assume that the cloud has small optical depth,
$\tau_\mathrm{cl}\ll1$, and it is ejected along $z$-axis, with bulk
velocity $\beta(z)\gg0$ directed approximately toward an observer
(inclination angle of the observer is denoted $\theta_{\rm{}o}$). 
Clearly, $\gamma\gg1$ implies the scattered photons are boosted in the
direction of motion. Although light bending increases the apparent size
of the star on the particle local sky, general relativity effects are
quite negligible on scattered photons moving straight away from the
center. Only few photons are scattered backwards, and therefore the
direct image greatly dominates the signal received by an observer.

\begin{figure*}
\includegraphics[width=\textwidth]{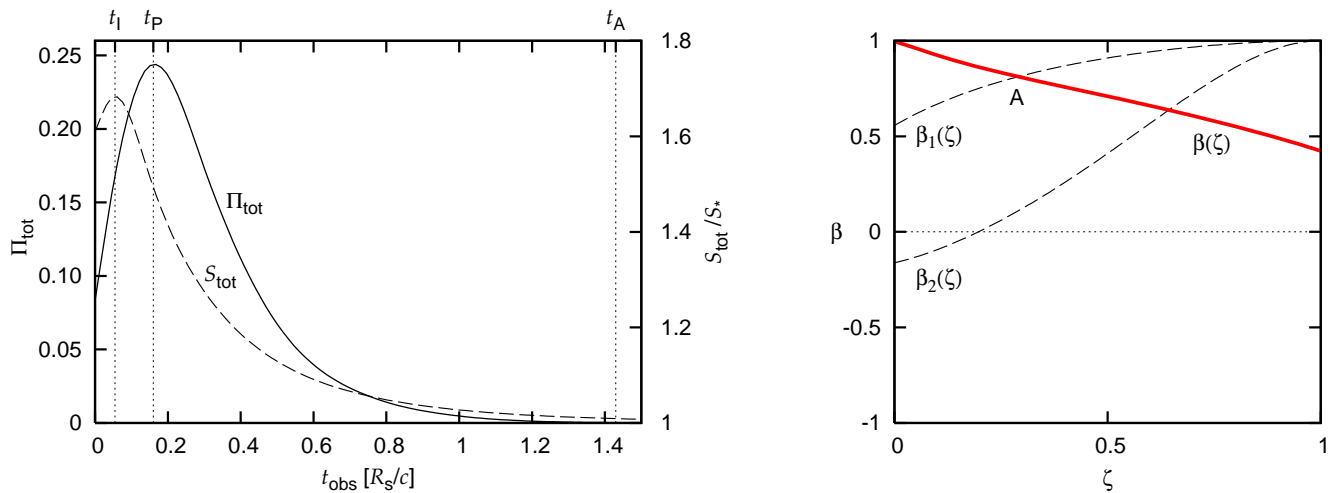}
\caption{Left: the variation of the total normalized radiation flux 
$S_\mathrm{tot}(t)$ and the corresponding degree of polarization
$\Pi_\mathrm{tot}(t)$. The case of ultra-relativistic ejection starting 
with initial $\gamma(R_\star)=10$. Right: The corresponding velocity
profile $\beta(\zeta)$ and critical velocities $\beta_{1,2}(\zeta)$ are
shown. Parameters of the plot are $\theta=17$~deg, $R_\star=1.2\rs$.
Polarization vanishes at the point \textsf{A} when
$\beta(\zeta)=\beta_1(\zeta)$; the corresponding time is
$t=t_\mathrm{A}$ in the left panel. Polarization vanishes once again at 
a later time, when $\beta(\zeta)=\beta_2(\zeta)$.}
\label{fig:frac}
\end{figure*}

The measured radiation flux $S_\mathrm{tot}$ has two components:  the
primary unpolarized flux $S_\star$ and the flux of partially polarized
radiation scattered in the cloud $S_\mathrm{cl}$. Their ratio can be
given in terms of redshifted intensities $\tilde{I}_\star$ and
$\tilde{I}_\mathrm{cl}$ of the star and of the cloud, and by the ratio
of solid angles occupied by the cloud and by the star on observer sky.
This provides us with the estimation of the expected fractional
polarization of the total signal. The ratio of fluxes is
\begin{equation}
s\equiv\frac{S_\mathrm{cl}}{S_\star} = \left(\frac{R_\mathrm{cl}}{R_\star}\right)^2
\frac{\tilde{I}_\mathrm{cl}}{\tilde{I}_\star}.
\label{eq:s}
\end{equation}
Intensity arriving from the cloud
$\tilde{I}_\mathrm{cl}=\xi^2(r)I_\mathrm{sc}$, where $I_\mathrm{sc}(r)$
is the locally emitted intensity, as given by eq.~(\ref{eq:isc}). The
scattered component is polarized with the magnitude $\Pi_\mathrm{cl}$,
as derived in the previous section. The total flux and the total
polarization are
\begin{equation}
S_\mathrm{tot}=(1+s)S_\star,\quad
\Pi_\mathrm{tot}=\frac{s\Pi_\mathrm{cl}}{1+s}.
\end{equation}

Substituting equation (\ref{eq:isc}) into (\ref{eq:s}) one can verify
that the flux ratio $s$ does not depend on the star intensity $I_\star$, 
and hence
\begin{equation}
s=\kappa f(\beta,r,\vartheta),
\end{equation}
where $\kappa\,\equiv\,\tau_\mathrm{cl}(R_\mathrm{cl}/R_\star)^2$ depends on
the size and density of the cloud, and $f$ includes the geometry of the
radiation field and the beaming/aberration effects arising from the
cloud motion. We consider situations when $\kappa$ is small; then the two
contributions to the radiation intercepted by the observer become
comparable only in case of strong beaming, which leads to $f\gg1$ for
small observing angles.

The moment of observation, i.e.\ the arrival time of photons
$t_\mathrm{obs}(r\rightarrow\infty)$, is related to the moment of
emission $t$ by
\begin{equation}
t_\mathrm{obs} \simeq t-\frac{1}{c}\big[z(t)-z_0\big]\cos\theta,
\end{equation}
where we set $t(z_0)=0$ for the initial time and $t_\mathrm{obs}=0$ for
the moment when the signal arrives to the observer. Notice that this
estimate is sufficient for direct image photons discussed in this
subsection, but it would not be appropriate for higher-order image
photons in the next subsection (in Schwarzschild geometry one can express 
time of arrival in terms of elliptic integrals, proceeding in the same 
analytical manner as above in the calculation of the ray trajectory; see 
also \citealt{boz04}; \citealt{cad05}).

\begin{figure}
\includegraphics[width=0.5\textwidth]{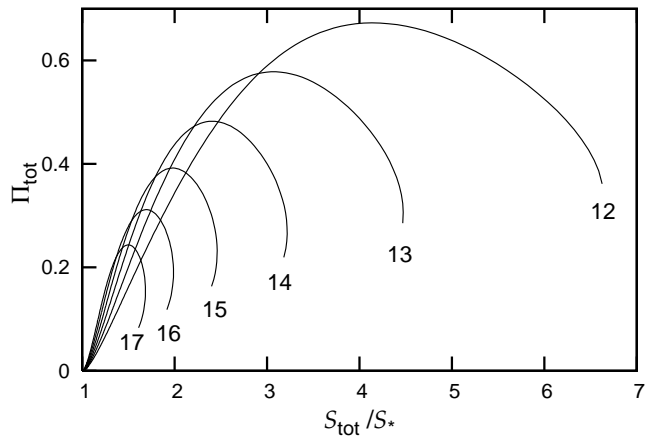}
\caption{The relation between the normalized
radiation flux and the total polarization. The magnitude of
polarization $\Pi_{\rm{}tot}$ reaches up to
$\sim65$\% for suitable view angles. Values of $\theta_{\rm{}o}$ 
(in degrees) are given with the curves. Other parameters are 
the same as in previous figure.}
\label{fig:frac3}
\end{figure}

The temporal behaviour is shown in figure~\ref{fig:frac}. We assume that
the cloud has been pre-accelerated to large initial speed $\beta(t=0)$
near the star surface. The graph captures the subsequent
phase of gravitational and radiative deceleration. The scattered light
contributes significantly to the total signal only for a short initial
phase (a peak occurs in the graph). The local maxima of the radiation
flux (at $t=t_\mathrm{I}$) and of polarization (at $t=t_\mathrm{P}$) can
be understood in terms of beaming: most of radiation from the cloud is
emitted in a cone with the opening angle $\sim1/\gamma$ about the
direction of motion. For small viewing angle
($\theta_{\rm{}o}\lesssim13$~deg) the observer was initially located
outside this cone but, as time goes, the electron decelerates, the cone
opens  up and the observer intercepts more radiation. The maximum
observed polarization occurs with a certain delay
$t_\mathrm{P}-t_\mathrm{I}$ (proportional to $M$) after the peak of
radiation flux. Subsequent decay of the signal is connected with a
diminished scattering power of the cloud and the overall dilution of the
radiation field. The observed polarization and the flux are lagged with
each other and sensitive to the angle of observation. This behaviour is
clearly seen also in figure~\ref{fig:frac3}, where we assumed several
different viewing angles.

We selected large initial velocity in this example, otherwise
the effects of aberration and the Doppler boosting would be less
prominent, time-scales longer, and the effect of fractional
polarization crossing zero point would disappear.
The time span of this plot can be scaled
according to the light-crossing time in physical units,
\begin{equation}
t\simeq1.5\,\frac{\rs}{c}=1.5\times10^{-4}\frac{M}{10M_\odot}\quad\mathrm{[sec]},
\end{equation}
i.e.\ proportionally to the central mass. The polarization magnitude is
correlated with the intensity (this correlation was already noticed for the
isotropic radiation in the right panel of Fig.~\ref{fig:iso}).

\subsection{A cloudlet at rest ({\mbox{\protect\boldmath{$\gamma=1$}}})} 
\label{sec:static}
An interplay between gravity and the ambient radiation stalls the bulk
motion, $\beta(t)\rightarrow0$. Scattered photons are then no 
longer boosted in the outward direction, and so
the higher order (highly bent) rays can provide a non-negligible
contribution to the observed light after encircling the star. This of
course requires large compactness; we set $R_\star=\frac{3}{2}\rs$
hereafter. Again we assume an observer near $z$-axis and a cloudlet with
a small size, $\psi\ll1$. Unlike a more traditional application of the 
lense geometry, the cloudlet is placed at an arbitrary finite distance
$z{\,\equiv\,}z(t)$ above the star and the deflection angle does not have
to be small.

Let us consider rays making a single round (by the angle
$\Theta=2\pi\pm\theta_{\rm{}o}$), with a radial turning point at 
pericenter $r=r_{\rm{}p}$ tightly above the photon circular orbit. As
mentioned in subsect.~\ref{sec:dynamics}, the equilibrium radius 
depends on the star luminosity. Once the cloudlet settles at the
equilibrium point $r=R_{\rm{}eq}$, scattered photons are no more boosted
to high energy and the collimation effect disappears. In this situation
relatively more light is backscattered in the direction  toward the
photon orbit. Part of these photons form a retrolensed image
\citep{hol02}, which may also reach the observer. We thus now calculate
(de)magnification of light also for the two first-order images, which
give the most significant contribution and may influence the net
polarization at infinity. To this aim we need to consider rays starting
near above the star, passing through pericenter and eventually escaping
to infinity (the retrolensing geometry; see \citealt{oha87};
\citealt{vir00}; \citealt{boz02}, and references cited therein). 

\begin{figure*}
\includegraphics[width=0.49\textwidth]{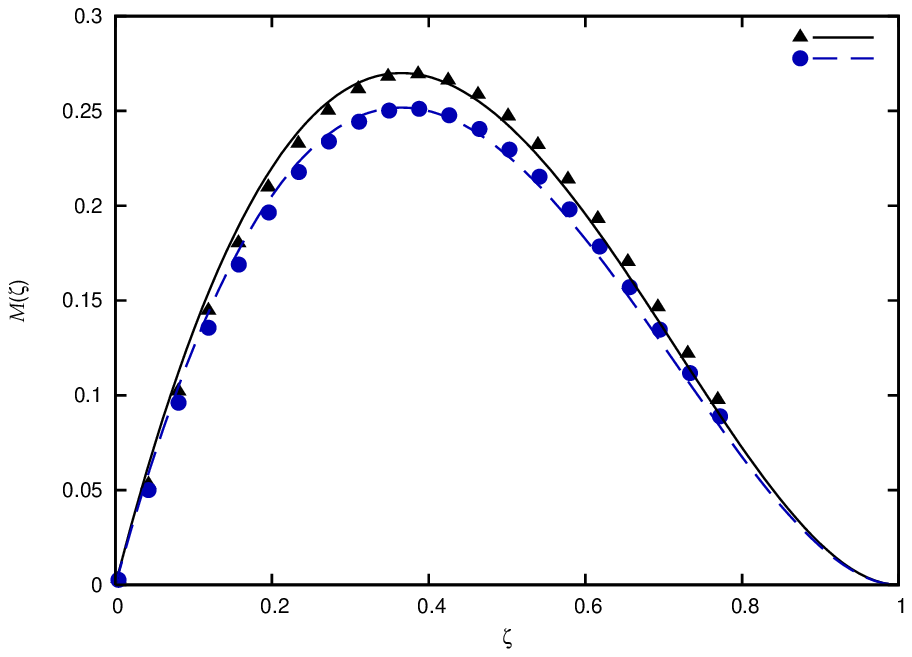}
\hfill
\includegraphics[width=0.49\textwidth]{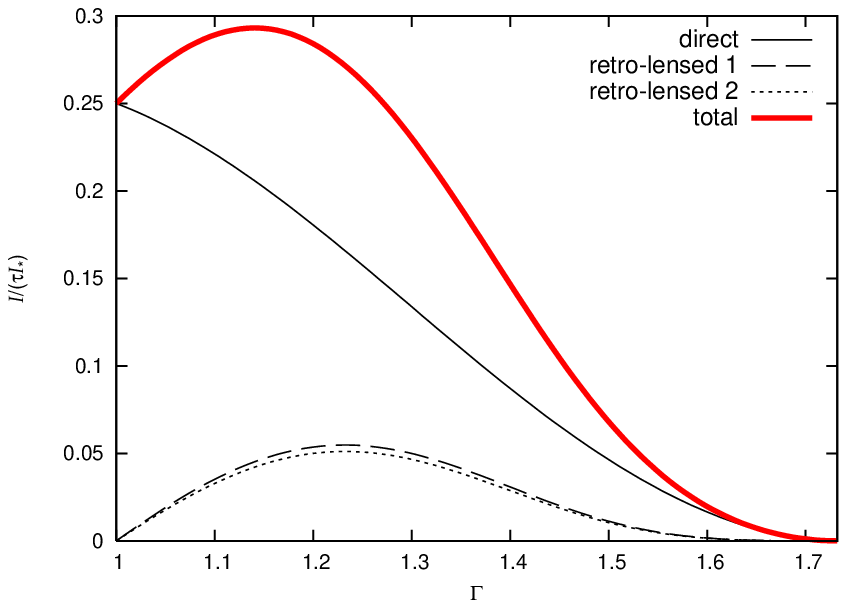}
\caption{Left: 
the gain factor $\mathcal{M}(\Theta,\zeta)$ according to the
approximation formula (\ref{eq:mu}). Two curves are shown as a function
of the source distance $\zeta$, for $\Theta=358\dg$ (solid line) and
$\Theta=362\dg$ (dashed line). For comparison, exact (numerically
computed) values are also plotted with triangles and circles. Right:
normalized intensity of the direct image (thin solid line) and the two
retrolensed images  (dashed and dotted lines) of light scattered from a
particle residing in the equilibrium point $z=R_{\rm{}eq}$, as a
function of the Eddington parameter. Notice that the contribution
of the two retrolensed images is almost identical and it amounts up to
$\sim20$\% of the total signal (thick solid line). Inclination
$\theta_{\rm{}o}=2\dg$ in both panels.}
\label{fig:mu}
\end{figure*}

Two arcs are formed which merge together in the Einstein ring (with
radius just above $b_{\rm{}c}=\frac{3}{2}\,\sqrt{3}\,\rs$) if the
observer is aligned with the source. By integrating null geodesics,
expanding the elliptic integrals near the pericenter
$r_{\rm{}p}\sim\frac{3}{2}\rs$, assuming the deflection angle close to
$\Theta\sim2\pi$, and keeping only the leading terms we obtain the
desired width of the two retrolensing images,
\begin{equation}
\delta{b}(\phi)=2\,\delta{b_0}\,\left(\psi^2-\theta^2_{\rm{}o}\sin^2\phi\right)^{1/2}
\end{equation}
where $\delta{b_0}=K(z)\,e^{-2\pi}$, $|\phi|\leq\arcsin(\bar{\psi}/\theta_{o})$,
$\bar{\psi}=\mbox{Min}\{\theta_{\rm{}o},\psi\}$, and
\begin{equation}
K=\frac{6^3\,3\,\sqrt{3}}{2}\,\frac{\sqrt{3}-1}{\sqrt{3}+1}\,
\frac{\sqrt{3}-\sqrt{1+3u}}{\sqrt{3}+\sqrt{1+3u}},\quad
u(z)\equiv\frac{\rs}{z}.
\end{equation}
We remark that $\Theta\sim2\pi$ was assumed for simplicity only.
The case of arbitrary $\Theta$ can be treated in similar way.

Time dependency of the arcs is caused by the scatterer motion, $z=z(t)$.
We integrate over the cross-section of the arc images to derive their
total luminosity and polarization at each time moment. Higher-order
images suffer from the de-magnifying influence of the light bending,
which reduces their luminosity, unless a special geometrical alignment
of the source and the observer occurs and favours the opposite effect of
a caustic. This can be quantified by the gain factor, ${\mathcal{M}}$,
which determines the ratio of fluxes received in retrolensed/direct
images. The problem translates to evaluating the ratio of solid angles,
$\mathcal{M}\,\equiv\,\frac{{\rm{}d}\Omega_{\rm{}i}}{{\rm{}d}\Omega_{\rm{}o}}$,
where indices ``i/o'' refer to the angular size of the source
with/without taking the light bending into account. In the case of a
small (but finite) size cloudlet, we find
\begin{equation}
{\mathcal{M}}(z)=6\,\sqrt{3}\,K(z)\,e^{-2\pi}\psi^{-1}\,
\Lambda(\theta_{\rm{}o}/\psi),
\end{equation}
where the term 
\begin{equation}
\Lambda(k)\,\equiv\,\frac{2}{\pi}\,E(\Phi,k),\quad
\Phi\,\equiv\,\arcsin\left[\mbox{Min}\left\{k^{-1},1\right\}\right],
\label{eq:phim}
\end{equation}
arises from the integration over the Einstein arcs.

For $\psi\ll\theta_{\rm{}o}$ the gain function is
\begin{equation}
\mathcal{M}(z,\Theta)\simeq{\textstyle{\frac{3}{2}}}\,\sqrt{3}\,K(z)\,
\frac{\rs^2}{z^2}\;\frac{\exp(-\Theta)}{|\sin\Theta|}.
\label{eq:mu}
\end{equation}
Formula (\ref{eq:mu}) reduces to eq.~(21) of \citet{oha87} for
$z\rightarrow\infty$, $\zeta(z)\rightarrow1$. In our situation,
eq.~(\ref{eq:mu}) requires that the cloudlet is sufficiently small in size and 
its motion is directed somewhat sideways with respect to the observer 
view angle. Figure \ref{fig:mu} compares ${\mathcal{M}}(\zeta,\Theta)$
with the corresponding result of a numerical integration, showing that
the approximation is sufficiently accurate for our purposes.

\begin{figure*}
\includegraphics[width=0.49\textwidth]{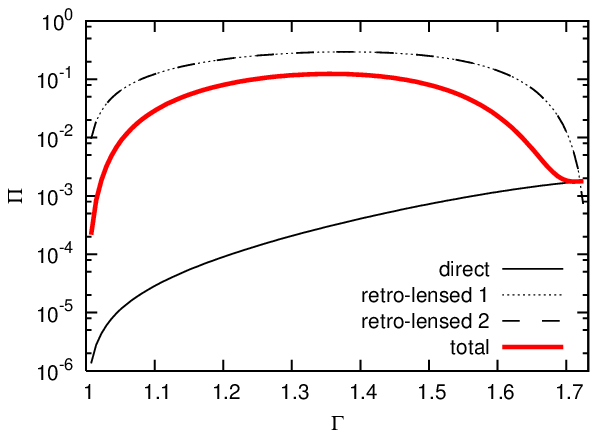}
\hfill
\includegraphics[width=0.49\textwidth]{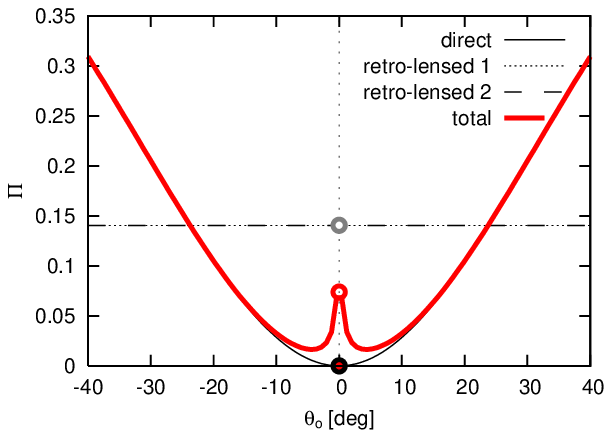}
\caption{Polarization magnitude from scattering on a particle at rest at 
$z=R_{\rm{}eq}$. This represents a cloudlet of angular radius  $\psi$ on
the local sky of the star. Left: the observed polarization magnitude as
a function of Eddington parameter $\Gamma$. Notice that $R_{\rm{eq}}$ is
a function of $\Gamma$, and so the graph covers the whole range of radii
from the star surface to infinity. The observer inclination is
$\theta_{\rm{}o}=2\dg$. Right: the corresponding polarization magnitude
for different inclinations and constant $\Gamma=1.6$ (in case of precise
alignment, $\theta_{\rm{}o}=0\dg$, polarization vanishes because of
symmetry).}
\label{fig:static}
\end{figure*}

\begin{figure*}
\includegraphics[width=0.4\textwidth]{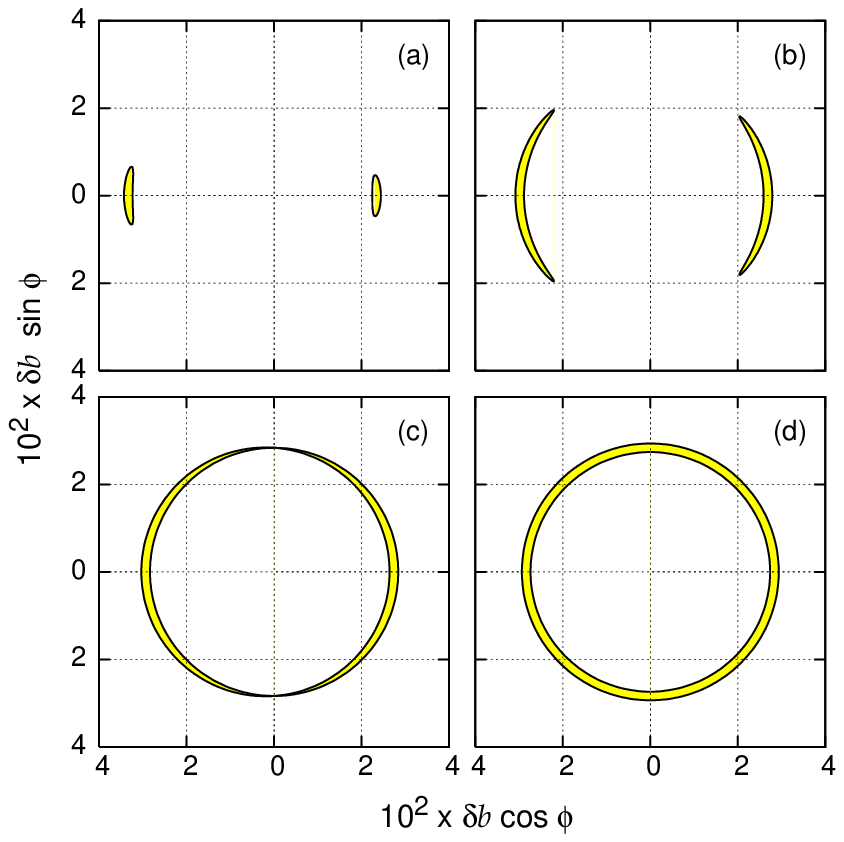}
\hfill
\includegraphics[width=0.58\textwidth]{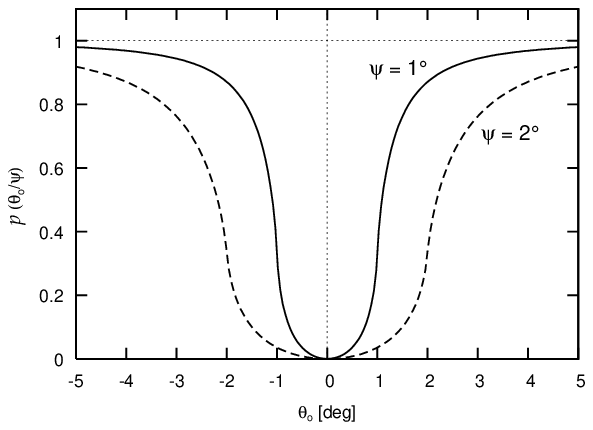}
\caption{Left: the form of Einstein arcs (a--c) and the ring (d) 
corresponding to the  retrolensing images in polar coordinates
$(b,\phi)$ in the observer plane. The source is supposed to be a
circular target of angular radius $\psi=2\dg$, located on $z$-axis  at
distance $z=3\rs$. The observer is at $r\rightarrow\infty$ and she has a
small angular offset from the perfect alignment --
(a)~$\theta_{\rm{}o}=10\dg$, (b)~$\theta_{\rm{}o}=2\dg$, 
(c)~$\theta_{\rm{}o}=1\dg$, (d)~$\theta_{\rm{}o}=0\dg$.  Right: a
contribution to the polarization produced by the Einstein arcs. A detail
of the normalized magnitude $p$ is plotted for small values of
inclination; see eq.~(\ref{eq:piret}) for the definition of function
$p(\theta_{\rm{}o}/\psi)$. For large $\theta_{\rm{}o}$ the magnitude of
polarization saturates at roughly constant level, equal to the
polarization scattered in the direction toward the photon circular
orbit. Two curves are parametrized by the angular size $\psi$ of the
cloudlet, as indicated in the plot.}
\label{fig:static2}
\end{figure*}

Adding the contributions from different parts of the source has a
depolarizing effect on the final signal, which we illustrate  in
figure~\ref{fig:static}. For the total magnitude of polarization of the
retrolensing images we find
\begin{equation}
\Pi_{\rm{}ret}=\Pi(\vartheta_{\rm{}ph})\,p(\theta_{\rm{}o}/\psi),
\label{eq:piret}
\end{equation}
where $\Pi(\vartheta_{\rm{}ph})$ is the polarization magnitude of
light scattered in the direction towards the photon circular orbit
and
\begin{equation}
p(k)\,\equiv\,\frac{2}{\pi\Lambda(k)}\int_0^{\Phi}\cos2\phi\,
\sqrt{1-k^2\sin^2\phi}\;{\rm{}d}\phi.
\end{equation}
Functions $\Lambda$ and $\Phi$ were defined in eq.~(\ref{eq:phim}).
Function $p$ determines the shape of retrolensing images in the observer
plane; see figure~\ref{fig:static2} (we resolve a narrow trace of these
arcs on observer sky by enlarging their separation
$\delta{b}\,\equiv\,b-b_{\rm{}c}$ from the critical radius
$b_{\rm{}c}$). Polarization vectors of all three images have the same
orientation, but they experience different time delays and lensing along
each trajectory. The contribution of the retrolensed images is now
evident, and quite significant. Notice that the angle
$\vartheta_{\rm{}ph}$ is the apparent angular size of the photon orbit
as seen on the local sky of the cloudlet. It enters in
eq.~(\ref{eq:piret}) because higher-order images are formed almost
exclusively by light scattered on the photon circular orbit. In our
case, $\vartheta_{\rm{}ph}=\alpha_\star(z)$. The polarization magnitude 
drops sharply if the observer inclination is less than the angular size
of the cloudlet.

\begin{figure*}
\includegraphics[width=0.49\textwidth]{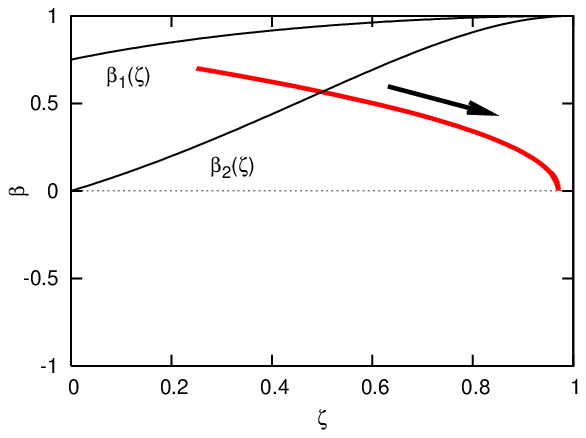}
\hfill
\includegraphics[width=0.49\textwidth]{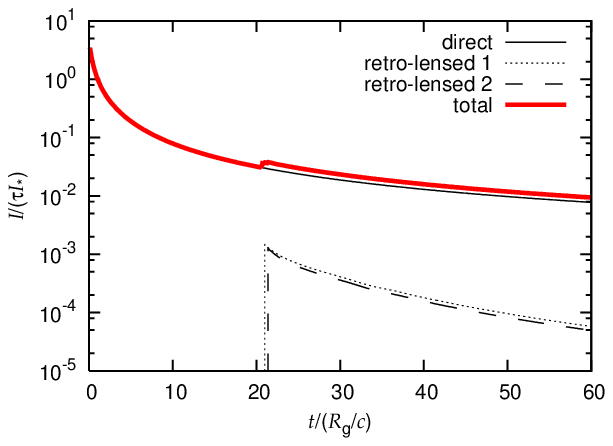}
\caption{An outward-moving particle
decelerating in the gravitational and weak radiation fields. Left: the
trajectory in the ($\beta,\zeta$)-plane, i.e.\ dimensionless velocity
versus distance (thick solid line). Velocity changes in the direction
of an arrow. Right: the radiation flux of scattered light as a function
of time for the direct and two retrolensed images. The initial condition
is $\beta=0.7$ at $r=2\rs$. The observer inclination is
$\theta_{\rm{}o}=5\dg$.}
\label{fig:ael1}
\end{figure*}

\begin{figure*}
\includegraphics[width=0.49\textwidth]{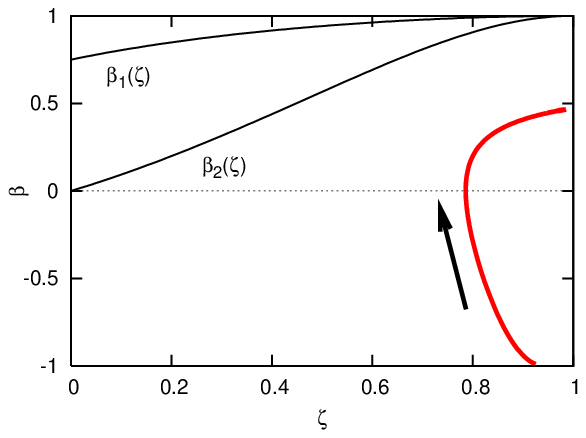}
\hfill
\includegraphics[width=0.49\textwidth]{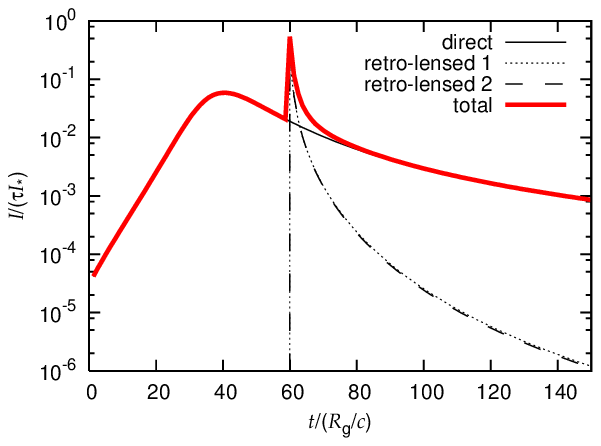}
\caption{The same as in Fig.~\ref{fig:ael1}, but for fast inward
motion of the scatterer and strong radiation of the star ($\Gamma=10$).
We set $\beta=-0.99$, $r=20\rs$ as an initial condition at $t=0$. The
signal of the higher-order image flashes for a brief moment around
dimensionless time $t\simeq60$. The intense radiation of the star
reverses the particle velocity to the outward motion at later stages.
The observer inclination is $\theta_{\rm{}o}=2\dg$.}
\label{fig:ael2}
\end{figure*}

\begin{figure*}
\includegraphics[width=0.49\textwidth]{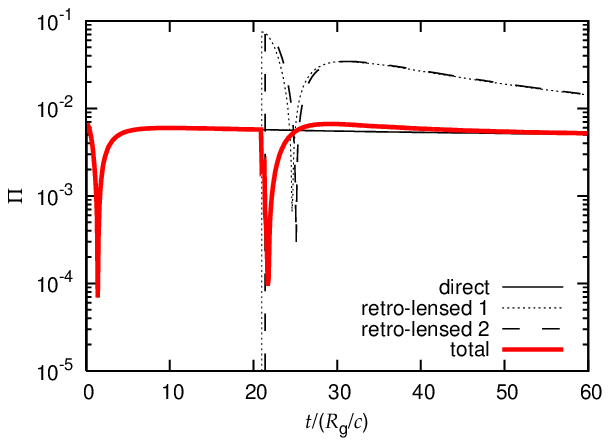}
\hfill
\includegraphics[width=0.49\textwidth]{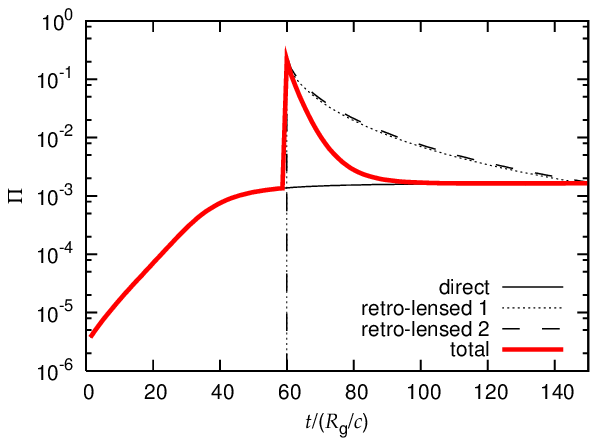}
\caption{The magnitude of
polarization corresponding to the case shown in Fig.~\ref{fig:ael1} 
(left panel) and in Fig.~\ref{fig:ael2} (right panel). The retrolensed
signals are delayed with respect to the direct signal. The delay has a
value  characteristic to light-travel time along the  photon circular
orbit (it scales proportionally to the mass of the central body).
The total polarization is suppressed or enhanced depending on the mutual
relation between the polarization of direct and retrolensed photons.}
\label{fig:ael3}
\end{figure*}

\subsection{Comparison between an outflow and an inflow}
\label{sec:inward}
Now we consider an intermediate situation with moderate velocity of the 
bulk motion (both an outflow or an inflow, i.e.\ $\beta\gtrless0$). For
a moderate outflow velocity, the result is shown in
figure~\ref{fig:ael1}. We consider a particle on the decelerating branch
of the trajectory in a weak radiation field, $\Gamma\rightarrow0$, which
eventually reaches the turning point (and starts falling afterwards).
The two retrolensed images contribute about $10$\% of the scattered flux
at maximum and the trajectory crosses $\beta=\beta_2(\xi)$ curve, where
the polarization vector swings its direction. The outcome is quite
different for matter infalling onto the star, because  scattered photons
are boosted in the downward direction and a considerable amount of light
is then directed on the photon orbit. As a result, the retrolensed
images are more pronounced and they cause a brief flash of light. The
resulting signal is shown in figure~\ref{fig:ael2}.

The effect of retrolensed images is clearly visible in the polarization
curve; see figure~\ref{fig:ael3}. The case shown in the left panel
exhibits a brief drop of the polarization magnitude when velocity
crosses $\beta=\beta_2(\zeta)$ (the direct image arrives at $t\sim1$ in
dimensionless units). At this moment the polarization changes its
orientation between transversal and the longitudinal one. Then the
signal restores back to a non-zero value and the same behaviour repeats
when the retrolensed image arrives after a certain delay (at $t\sim21$).
The case shown in the right panel exhibits a similar flash, also caused
by the contribution of the retrolensed photons. However, now we observe
a single fluctuation, which is actually an increase of the polarization
magnitude; this is because the case shown here corresponds to
transversal polarization during the whole observation and the trajectory
does not cross any of the critical curves $\beta=\beta_{1,2}$.

\section{Conclusions}
Our calculation here is self-consistent in the sense that the motion of
the blob and of photons, and the resulting polarization are mutually
connected. We concentrated ourselves on gravitational effects and
neglected other intervening processes, first of all the effect of
magnetic fields to which polarization is sensitive (see e.g.\
\citealt{ago96}). This allowed us to compare polarization magnitudes
of direct and retrolensed images, which could point to the presence of a
highly compact body. We have noticed the mutual delay between the signal
formed by photons of different order. The delay is characteristic to the
effect and has a value proportional to the central mass.

Polarimetric properties are susceptible to large changes depending on
detailed physics and geometry of the source, and this is not only the
advantage, which could help us to trace how different objects are
functioning, but also a complication. In particular, the polarization is
sensitive to the source orientation and its magnitude fluctuates from
case to case. Our results are useful for testing more complex and
astrophysically realistic models with up-scattering of soft photons in
jets and fast flows around black holes. In the black hole case the
primary photons would be provided by an accretion disc rather than the
star surface and, hence, one can no more take advantage of its spherical
symmetry, which helped us to simplify our calculations here.
Nonetheless, the same formalism can be employed and similar features of
time-dependent polarization lightcurves are expected; strong gravity
plays a vital role again. Putting this in another way: provided a
`realistic' equation of state implies neutron star radii greater than
the photon circular orbit, detection of the signatures of retrolensing
images, which we discussed above, would exclude a neutron star as a
candidate on the central body.

\section*{Acknowledgments}
The authors thank John Miller for his advice on the current status of
neutron star modelling, and an anonymous referee for critical comments
on the first version of the paper. VK appreciates fruitful discussions
with participants of the Aspen Center for Physics 2005 workshop
Revealing Black Holes. The financial support for this research has been
provided by the Academy of Sciences (grant IAA\,300030510) and by
the Czech Science Foundation (grant 205/03/0902). The Astronomical
Institute has been operated under the project AV0Z10030501.

\label{lastpage}


\begin{thebibliography}{99}
\bibitem[\protect\citeauthoryear{Abramowicz et al.}{1990}]{abr90}Abramowicz M.~A., Ellis G.~F.~R., Lanza~A., 1990, ApJ, 361, 470
\bibitem[\protect\citeauthoryear{Agol \& Blaes}{1996}]{ago96}Agol~E., Blaes~O., 1996, MNRAS, 282, 965
\bibitem[\protect\citeauthoryear{Alcock et al.}{1986}]{alc86}Alcock~C., Farhi~E., Olinto~A., 1986, ApJ, 310, 261
\bibitem[\protect\citeauthoryear{Angel}{1969}]{ang69}Angel J.~R.~P., 1969, MNRAS, 158, 219
\bibitem[\protect\citeauthoryear{Bao et al.}{1997}]{bao97}Bao~G., Hadrava~P., Wiita P.~J., Xiong~Y., 1997, ApJ, 487, 142 
\bibitem[\protect\citeauthoryear{Begelman \& Sikora}{1987}]{beg87}Begelman M.~C., Sikora~M., 1987, ApJ, 322, 650
\bibitem[\protect\citeauthoryear{Beloborodov}{1998}]{bel98}Beloborodov A.~M., 1998, ApJ, 496, L105
\bibitem[\protect\citeauthoryear{Bonometto et al.}{1970}]{bon70}Bonometto~S., Cazzola~P., Saggion~A., 1970, A\&A, 7, 292
\bibitem[\protect\citeauthoryear{Born \& Wolf}{1964}]{bor64}Born~M., Wolf~E., 1964, Principles of Optics (Pergamon Press, Oxford)
\bibitem[\protect\citeauthoryear{Bower et al.}{2005}]{bow05}Bower G.~C., Falcke~H., Wright M.~C.~H., Backer D.~C., 2005, ApJ, 618, L29
\bibitem[\protect\citeauthoryear{Bozza}{2002}]{boz02}Bozza~V., 2002, Phys. Rev. D, 66, 103001
\bibitem[\protect\citeauthoryear{Bozza \& Mancini}{2004}]{boz04}Bozza~V., Mancini~L., 2004, Gen. Rel. Grav., 36, 435
\bibitem[\protect\citeauthoryear{Brown et al.}{1998}]{bro98}Brown E.~F., Bildsten~L., Rutlege R.~E., 1998, ApJ, 504, L95
\bibitem[\protect\citeauthoryear{Brown \& McLean}{1977}]{bro77}Brown J.~C., McLean I.~S., 1977, A\&A, 57, 141
\bibitem[\protect\citeauthoryear{\v{C}ade\v{z} \& Kosti\'c}{2005}]{cad05}\v{C}ade\v{z}~A., Kosti\'c~U., 2005, submitted (gr-qc/0405037)
\bibitem[\protect\citeauthoryear{Celotti \& Matt}{1994}]{cel94}Celotti~A., Matt~G., 1994, MNRAS, 268, 451
\bibitem[\protect\citeauthoryear{Chandrasekhar}{1960}]{cha60}Chandrasekhar~S., 1960, Radiative Transfer (Dover Publ., New York)
\bibitem[\protect\citeauthoryear{Chandrasekhar}{1992}]{cha92}Chandrasekhar~S., 1992, The Mathematical Theory of Black Holes (Oxford Univ. Press, New York)
\bibitem[\protect\citeauthoryear{Chandrasekhar \& Ferrari}{1991}]{cha91}Chandrasekhar~S., Ferrari~V., 1991, Proc. R. Soc. Lond. A, 434, 449
\bibitem[\protect\citeauthoryear{Cocke \& Holm}{1972}]{coc72}Cocke~W.~J., Holm D.~A., 1972, Nature Physical Science, 240, 161
\bibitem[\protect\citeauthoryear{Combi et al.}{2004}]{com04}Combi J.~A., Cellone S.~A., Mart\'{\i}~J., Rib\'o~M., Mirabel I.~F., Casares~J., 2004, A\&A, 427, 959
\bibitem[\protect\citeauthoryear{Connors et al.}{1980}]{con80}Connors P.~A., Stark R.~F., Piran~T., 1980, ApJ, 235, 224
\bibitem[\protect\citeauthoryear{Dey et al.}{1998}]{dey98}Dey~M., Bombaci~I., Dey~J., Ray~S., Samanta B.~C., 1998, Physics Letters B, 438, 123
\bibitem[\protect\citeauthoryear{Fox}{1994}]{fox94}Fox G.~K., 1994, ApJ, 435, 372
\bibitem[\protect\citeauthoryear{Fukue \& Hachiya}{1999}]{fuk99}Fukue~J., Hachiya~M., 1999, PASJ, 51, 185
\bibitem[\protect\citeauthoryear{Ghisellini et al.}{2004}]{ghi04}Ghisellini~G., Haardt~F., Matt~G., 2004, A\&A, 413, 535
\bibitem[\protect\citeauthoryear{Ghisellini \& Lazzati}{1999}]{ghi99}Ghisellini~G., Lazzati~D., 1999, MNRAS, 309, L7
\bibitem[\protect\citeauthoryear{Haensel}{2003}]{hae03}Haensel~P., 2003, in Final Stages of Stellar Evolution, eds. C.~Motch \& J.-M. Hameury (EAS Publications Series, EDP Sciences), p.~249
\bibitem[\protect\citeauthoryear{Haensel \& Zdunik}{1990}]{hae90}Haensel~P., Zdunik J.~L., 1990, A\&A, 227, 431
\bibitem[\protect\citeauthoryear{Holz \& Wheeler}{2002}]{hol02}Holz D.~E., Wheeler J.~A., 2002, ApJ, 578, 330
\bibitem[\protect\citeauthoryear{Hor\'ak}{2005}]{hor05}Hor\'ak~J., 2005, Thesis, submitted (Charles University Prague)
\bibitem[\protect\citeauthoryear{Keane et al.}{2001}]{kea01}Keane A.~J., Barrett R.~K., Simmons J.~F.~L., 2001, MNRAS, 321, 661
\bibitem[\protect\citeauthoryear{Kokkotas et al.}{2004}]{kok04}Kokkotas K.~D., Ruoff~J., Andersson~N., 2004, Phys. Rev. D, 70, 043003
\bibitem[\protect\citeauthoryear{Lattimer \& Prakash}{2001}]{lat01}Lattimer J.~M., Prakash~M., 2001, ApJ, 550, 426
\bibitem[\protect\citeauthoryear{Lattimer \& Prakash}{2004}]{lat04}Lattimer J.~M., Prakash~M., 2004, Science, 304, 536
\bibitem[\protect\citeauthoryear{Lazzati et al.}{2004}]{laz04}Lazzati~D., Rossi~E., Ghisellini~G., Rees M.~J., 2004, MNRAS, 347, L1
\bibitem[\protect\citeauthoryear{Levinson \& Eichler}{2004}]{lev04}Levinson~A., Eichler~D., 2004, ApJ, 613, 1079
\bibitem[\protect\citeauthoryear{Lightman \& Shapiro}{1975}]{lig75}Lightman A.~P., Shapiro S.~L., 1975, ApJ, 198, L73
\bibitem[\protect\citeauthoryear{Mazur \& Mottola}{2004}]{maz04}Mazur~E., Mottola P.~O., 2004, Proc. Nat. Acad. Sci. 111, 9545 (gr-qc/0109035)
\bibitem[\protect\citeauthoryear{McClintock et al.}{2004}]{mcc04}McClintock J.~E., Narayan~R., Rybicki G.~B., 2004, ApJ, 615, 402
\bibitem[\protect\citeauthoryear{Melia \& K\"onigl}{1989}]{mel89}Melia~F., K\"onigl~A., 1989, ApJ, 340, 162
\bibitem[\protect\citeauthoryear{Miller et al.}{1998}]{mil98}Miller J.~C., Shahbaz~T., Nolan L.~A., 1998, MNRAS, 294, L25
\bibitem[\protect\citeauthoryear{Ohanian}{1987}]{oha87}Ohanian H.~C., 1987, Am. J. Phys., 55, 428
\bibitem[\protect\citeauthoryear{Pineault}{1977}]{pin77}Pineault~S., 1977, MNRAS, 179, 691
\bibitem[\protect\citeauthoryear{Poeckert \& Marlborough}{1976}]{poe76}Poeckert R., Marlborough J.~M., 1976, ApJ, 206, 182
\bibitem[\protect\citeauthoryear{Portsmouth \& Bertschinger}{2005}]{por05}Portsmouth~J., Bertschinger~E., 2005, submitted (astro-ph/0412094)
\bibitem[\protect\citeauthoryear{Poutanen}{1994}]{pou94}Poutanen~J., 1994, ApJSS, 92, 607
\bibitem[\protect\citeauthoryear{Rees}{1976}]{ree75}Rees M.~J., 1975, MNRAS, 171, 457
\bibitem[\protect\citeauthoryear{Rudy}{1978}]{rud78}Rudy R.~J., 1978, PASP, 90, 688
\bibitem[\protect\citeauthoryear{Rybicki \& Lightman}{1979}]{ryb79}Rybicki G.~B., Lightman A.~P., 1979, Radiative Processes in Astrophysics (Wiley, New York)
\bibitem[\protect\citeauthoryear{Scaltriti et al.}{1997}]{sca97}Scaltriti~F., Bodo~G., Ghisellini~G., Gliozzi~M., Trussoni~E., 1997, A\&A, 325, L29
\bibitem[\protect\citeauthoryear{Shaviv \& Dar}{1995}]{sha95}Shaviv N.~J., Dar~A., 1995, ApJ, 447, 863
\bibitem[\protect\citeauthoryear{Sikora \& Wilson}{1981}]{sik81}Sikora~M., Wilson D.~B., 1981, MNRAS, 198, 529
\bibitem[\protect\citeauthoryear{Sunyaev \& Titarchuk}{1985}]{sun85}Sunyaev R.~A., Titarchuk L.~G., 1985, A\&A, 143, 374
\bibitem[\protect\citeauthoryear{Synge}{1967}]{syn67}Synge J.~L., 1967, MNRAS, 136, 195
\bibitem[\protect\citeauthoryear{Viironen \& Poutanen}{2004}]{vii04}Viironen~K., Poutanen~J., 2004, A\&A, 426, 985
\bibitem[\protect\citeauthoryear{Virbhadra \& Ellis}{2000}]{vir00}Virbhadra K.~S., Ellis G.~F.~R., 2000, Phys. Rev. D, 62, 084003
\bibitem[\protect\citeauthoryear{Vokrouhlick\'y \& Karas}{1991}]{vok91}Vokrouhlick\'y~D., Karas~V., 1991, A\&A, 252, 835
\bibitem[\protect\citeauthoryear{Weber}{2005}]{web05}Weber~F., 2005, Progress in Particle and Nuclear Physics, 2005, 54, 193
\bibitem[\protect\citeauthoryear{Winterberg \& Phillips}{1973}]{win73}Winterberg~F., Phillips W.~G., 1973, Phys. Rev. D, 8, 3329
\end{thebibliography}
\end{document}